\documentclass{aa}  
\usepackage{graphicx}
\usepackage{txfonts}
\usepackage{hyperref}
\usepackage{graphicx}
\usepackage{subcaption}
\usepackage{amsmath}
\usepackage{txfonts}
\usepackage{hyperref}
\usepackage[normalem]{ulem}
\usepackage[x11names]{xcolor}
\usepackage{academicons}
\usepackage[flushleft]{threeparttable}
\usepackage{booktabs,caption}
\definecolor{orcidlogocol}{HTML}{A6CE39}
\begin{document} 

\title{Multi-wavelength synthesis of a flux rope-trapped mini-prominence eruption and post-flare coronal rain}

 \author{Samrat Sen$^*$ \inst{1, 2, 3} \href{https://orcid.org/0000-0003-1546-381X}{\includegraphics[scale=0.05]{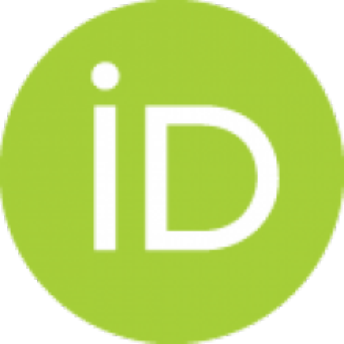}}, Alexander G. M. Pietrow\inst{4} \href{https://orcid.org/0000-0002-0484-7634}{\includegraphics[scale=0.05]{figures/orcid-ID.pdf}}, Veronika Jer{\v{c}}i{\'c}\inst{5} \href{https://orcid.org/0000-0003-1862-7904}{\includegraphics[scale=0.05]{figures/orcid-ID.pdf}}, Patrick Antolin \inst{6} \href{https://https://orcid.org/0000-0003-1529-4681}{\includegraphics[scale=0.05]{figures/orcid-ID.pdf}} 
   }

   \institute{
   Instituto de Astrof\'{i}sica de Canarias, 38205 La Laguna, Tenerife, Spain
   \and
   Universidad de La Laguna, 38206 La Laguna, Tenerife, Spain 
   \and
   Department of Physics, University of Helsinki, P.O. Box 64, FI-00014 Helsinki, Finland
    \\
   $^*$\email{samratseniitmadras@gmail.com; samrat.sen@helsinki.fi}
   \and
   Leibniz-Institut f\"ur Astrophysik Potsdam (AIP), An der Sternwarte 16, 14482 Potsdam, Germany
   \and
    NASA Goddard Space Flight Center, Greenbelt, MD, USA
    \and
    School of Engineering, Physics, and Mathematics, Northumbria University, Newcastle upon Tyne, NE1 8ST, UK
         }
         
   \date{Received: XXXX; accepted: XXXX}
   \date{}
 
  \abstract 
   {Small-scale eruptive phenomena in the solar corona, including miniature flux ropes and associated cool plasma condensations, are not fully understood despite increasing high-resolution observations. In particular, the phenomena including flux rope eruptions with trapped mini-prominence, and post-flare coronal rain for miniature loops has not been extensively explored within a unified magnetohdrodynamic (MHD) model.}
   {We aim to establish observational diagnostics of mini-prominence eruptions and post-flare coronal rain by bridging MHD simulations with synthetic observables across multiple wavelength regimes from extreme ultraviolet (EUV) to optical.}
   {We perform forward modeling based on a 2.5D resistive MHD simulation that captures homologous flux rope eruptions, in-situ condensation leading to a mini-prominence, and subsequent post-flare coronal rain. Synthetic diagnostics (imaging and spectra) are obtained using optically thin approximations for EUV and ultraviolet (UV) emissions, and non-local thermodynamic equilibrium (non-LTE) radiative transfer treatment for the H$\alpha$ line. Instrument resolution effects are incorporated to enable direct comparison with current observational capabilities.}
   {The synthetic EUV emission reveals the flux ropes as bright rim-like structures. The corresponding UV diagnostic shows bright region, which is co-spatial with the dark core due to embedded cool plasma ($\sim$ tens of kK) inside the flux rope, identifying a mini-prominence carried by the erupting flux rope. Spectral synthesis of Si IV 1402.77~\AA\ indicates an upward motion of the mini-filament, and reveals the presence of two predominant velocity components during eruption. At a later stage, thermal instability in post-flare arcades produces coronal rain with temperatures of $\approx 10^4$~K. The EUV diagnostics reveal brightening at the downstream of the rain blob, indicating localized heating associated with compressional effects. The H$\alpha$ spectral synthesis shows enhanced absorption signatures and red shifted profiles corresponding to down flows of the coronal rain blobs up to $\approx 23$ km s$^{-1}$, whereas the Si~IV~1402.77~\AA\ spectral profile shows the the maximum down flow velocity of $\approx50$ km~s$^{-1}$, highlighting the evidence of thermodynamic and kinematic structuring within the falling rain blobs.}
   {The synthetic diagnostics provide clear, multi-wavelength signatures that can guide future high-resolution observations, and highlight the importance of small-scale reconnection-driven processes in shaping the multi-thermal structure (between MK to kK) of the solar corona.}

   \keywords{Magnetic reconnection -- Magnetohydrodynamics (MHD) -- Methods: numerical -- Sun: corona -- Sun: filaments, prominences}

\titlerunning{Synthetic diagnostics of a mini-prominence eruption and post-flare coronal rain}
\authorrunning{Sen et al.}
\maketitle
%

\section{Introduction}\label{sec:intro}
Magnetic structures associated with solar eruptive events are often linked to magnetic flux ropes (MFRs). MFRs are characterized as bundles of magnetic field lines those are wrapped about a common axis \citep{2011:chen, 2014:priest-book, 2022:he}. They can become unstable and erupt under favorable conditions owing to ideal magnetohydrodynamic (MHD) instabilities, such as the kink and torus instabilities \citep{2007:fan, 2006:kliem}, or through resistive instability via magnetic reconnection \citep[][and references therein]{Antiochos:1999, Moore:2001, Sen-FMI:2025}. MFRs are sometimes capable of supporting cold ($\sim 10^4$ K) and dense ($\sim 10^{10}$ cm$^{-3}$) plasma clouds against gravity within their magnetic dips through magnetic tension. These cold-dense plasma clouds often appear as prominences when they are observed above the solar limb, and termed as filaments when observed at the disk. The miniature counter part of the large-scale filaments are called mini-filaments which has a length scale of $\lesssim 30$~Mm and widths of $\lesssim 2.2$~Mm \citep[][and references therein]{Wang-mini-filament:2000, Sterling_2015Natur.523..437S, Li2023_mini-prominence, Teng_mini-filament:2024}, whereas, mini-prominences can be regarded as the limb-view manifestations of mini-filaments. The mini-filaments can be supported by miniature flux ropes of comparable spatial scales, and can produce `mini coronal mass ejections' \citep{Innes:2009}, which are also associated with MFR eruptions. Therefore, mini-filament eruptions can often serve as proxies for the eruptions of small-scale flux ropes. Observational studies have reported that eruptions of mini-filaments can trigger other solar activities, such as flares \citep{Yang-mini-filament:2018} and coronal jets \citep{Hong:2011, Adams:2014, Pansesar:2016, 2016ApJ...821..100S}, which highlights the importance of these small-scale phenomena in driving larger scale solar activity. Modeling efforts by \cite{2017Natur.544..452W, 2018ApJ...852...98W} demonstrated the magnetic breakout mechanism \citep{Antiochos:1999} for large-scale coronal mass ejections, while explaining coronal jets involving mini-filaments.

The formation of the coronal cool-condensations in general are based on the similar underlying physical mechanisms due to imbalance between heating and cooling. Radiative losses play a central role in the formation of these structures. When radiative cooling locally exceeds the net heating rate, the temperature in that region can decrease to $\sim 10^4$ K, leading to the formation of cool, condensed plasma. Such condensations can arise in various magnetic configurations, for example in the vicinity of plasmoids (or flux ropes) within tearing-mediated current sheets \citep{Sen:2022, Sen:2023, DeJonghe:2025}, or in magnetic arcades \citep[][and references therein]{Veronica:2023}. These cool condensations may erupt together with flux ropes \citep[][and references therein]{zhao2017, Jenkins:2021} in the form of prominences (or filaments), or they may fall along magnetic field lines as coronal rain \citep[][and references therein]{2017A&A...603A..42X, 2021:wenzhi, 2022:xiaohong}. Coronal rain has been reported in post-flare loops (PFLs) within the framework of the classic loop-prominence model of \cite{1964:bruzek}. In this scenario, condensations form in PFLs as a consequence of flare heating \citep{1980:antiochos}, resulting in a single episode of coronal rain followed by its subsequent descent. This interpretation is supported by the observations of \cite{2014:scullion, 2016:scullion} and \cite{2015:liu}. More recently, several two- and three-dimensional models of purely coronal volumes have been developed \citep{2019A&A...624A..96C, 2020A&A...636A.112C, Sen:2024, Dion:2024, DeJonghe:2025}. In these models, the thermal non-equilibrium \citep[TNE;][]{1991:antiochos} scenario is excluded by construction, and instead the formation of localized cool-condensations are triggered by thermal instability \citep{1965ApJ...142..531F}.  

Inferring the magnetic topology of small-scale flux ropes, as well as the mechanisms driving their eruptions, remains challenging from observations alone. To address this limitation, we need to rely on numerical modeling. Modeling efforts aimed at capturing homologous eruptions with flux rope-trapped mini-prominence (or mini-filament), together with post-flare coronal rain within a single framework at such small spatial scales, have received limited attention to date. Recently, \cite{Sen:2024} (S24 hereafter) presented such a scenario using a resistive magnetohydrodynamic (MHD) simulation that captures these phenomena within a unified MHD framework that incorporates field-aligned thermal conduction, optically thin radiative losses, and steady (but spatially varying) background heating. This work demonstrates how magnetic reconnection facilitates the formation and eruption of homologous flux ropes, followed by the onset of coronal rain, without requiring any ad hoc prescription of localized heating. A parametric survey of this model further reveals that flux rope formation is favored in a case with higher magnetic shear (for a shearing angle $\approx 72^\circ$), whereas lower shear (shearing angle $\approx 26^\circ$) does not lead to flux rope formation and cool-condensations, as reported in \cite{Sen:2025b}. These findings suggest a potential link between eruptive processes and the formation of cool condensations in miniature coronal strands ambiance. Therefore, this model warrants further investigation through forward modeling to establish its observational counterparts and to identify the signatures of small-scale flux ropes, mini-prominences, and post-flare coronal rain, thereby providing guidance for future observations of such events. 

In the current work, we present the synthetic diagnostics (imaging and spectra) based on the S24 model. We perform forward modeling in optically thin approximation for ultraviolet (UV) and extreme-ultraviolet (EUV) regimes, and use non-local thermodynmic equilibrium (non-LTE) radiative transfer treatment in optical regime. The synthetic observables in simulation resolution, and a degraded resolution compatible to the existing observing facilities are intended to guide and assess future observational campaigns. These efforts aim to draw connection between eruptive processes, plasma thermodynamics, and radiative signatures in $\sim$~Mm-scale coronal ambiance, thereby establishing a bridge between MHD modeling and observations. 

The paper is organized as follows. In Section~\ref{sec:simulation}, we provide a brief overview of the underlying MHD model which is used as a basis of the synthesis carried out in this work. In Sections~\ref{sec:MFR-prominence} and \ref{sec:coronal-rain}, we present the main results of the synthetic observations, along with a description of the underlying physics. In Section~\ref{sec:summary}, we summarize the key findings, connect them with existing observations, and highlight the novelty of the work. Finally, in the conclusion, we discuss the caveats of the work, and outline the potential directions for future improvements. 

\begin{figure}[hbt!]
    \centering
    \includegraphics[width=1\linewidth]{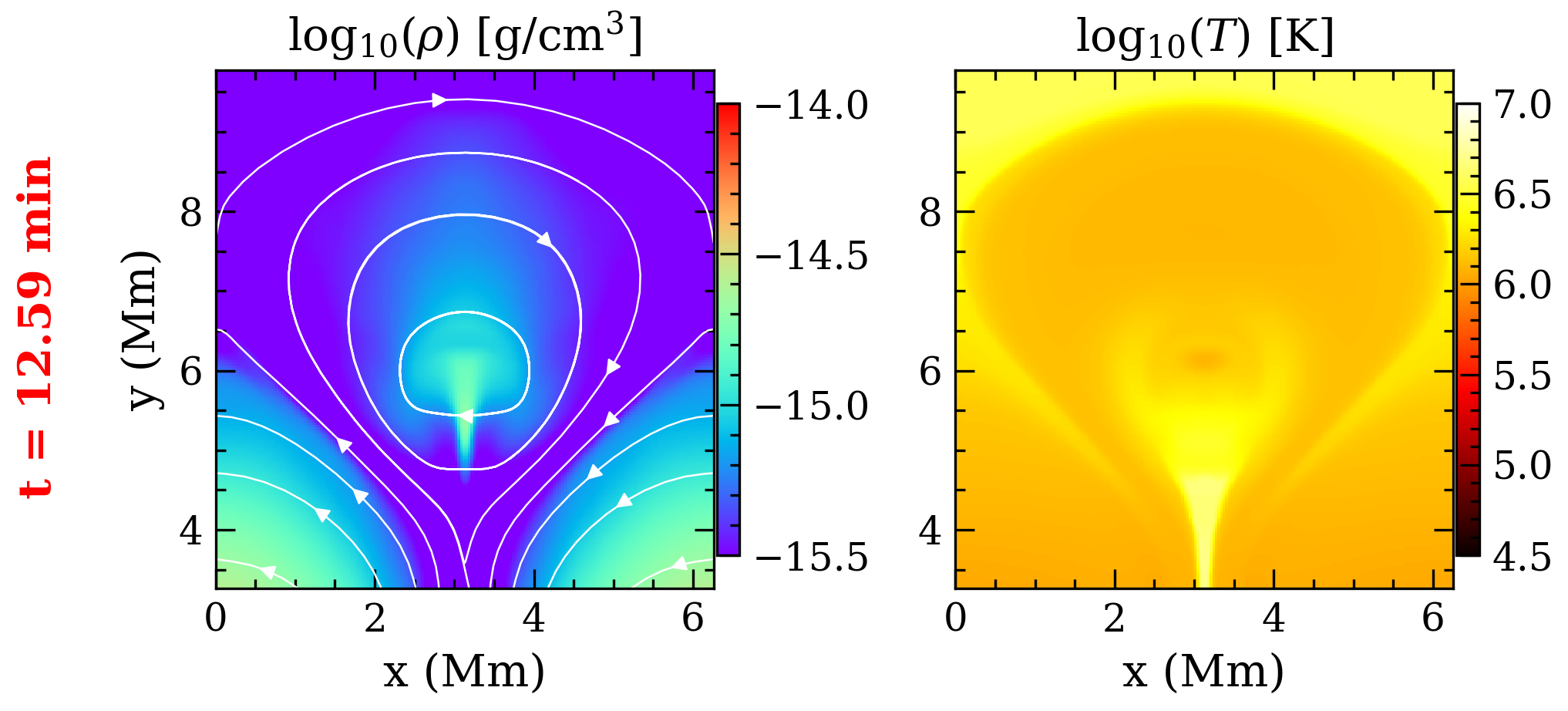}
    \includegraphics[width=1\linewidth]{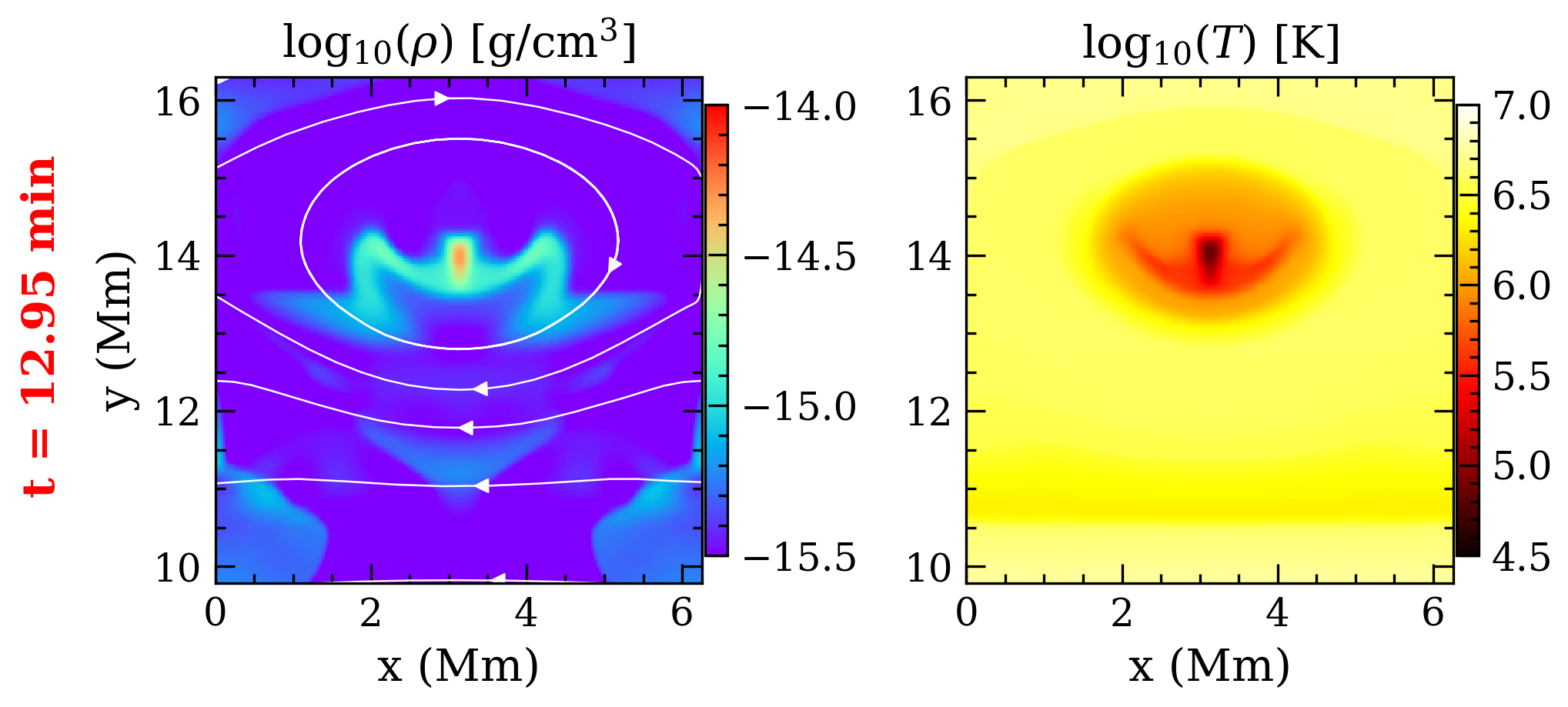}
    \includegraphics[width=1\linewidth]{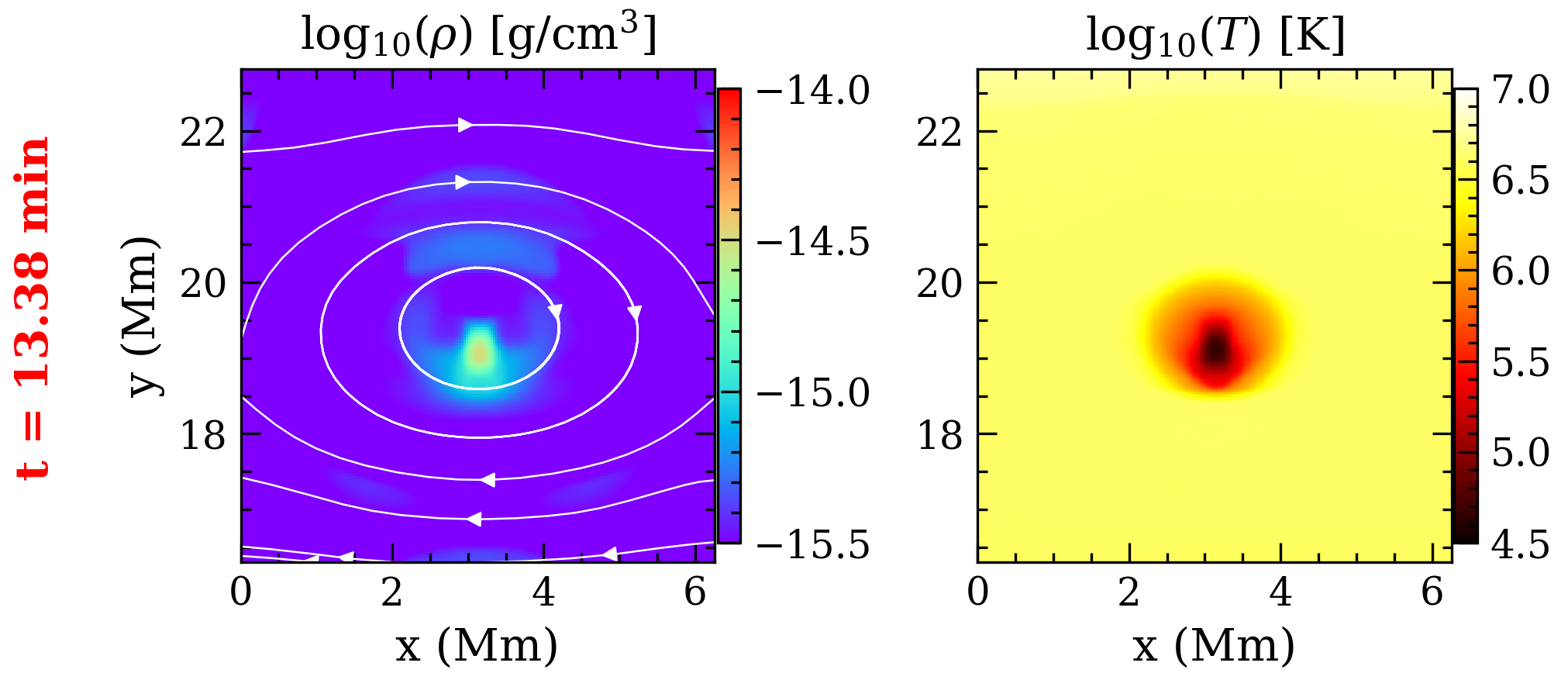}
    \caption{\textit{Top row:} Spatial distributions of the density ($\rho$), projected magnetic field lines (white) in the $x$–$y$ plane, and temperature ($T$) shown from left to right at $t=12.59$ min. The magnetic field lines indicate the presence of a flux rope between $y\approx 4$ and 9 Mm. \textit{Middle and bottom rows:} Same as the top row, but at the later times $t=12.95$~min and $t=13.38$~min respectively, illustrating the eruption of the flux rope. The temperature maps showcase the formation of localized cool structures which are co-spatial with the localized condensed structures trapped inside the flux rope as shown in the corresponding density maps.}
    \label{fig:MHD}
\end{figure}

\begin{figure}[hbt!]
    \centering
    \includegraphics[width=0.8\linewidth]{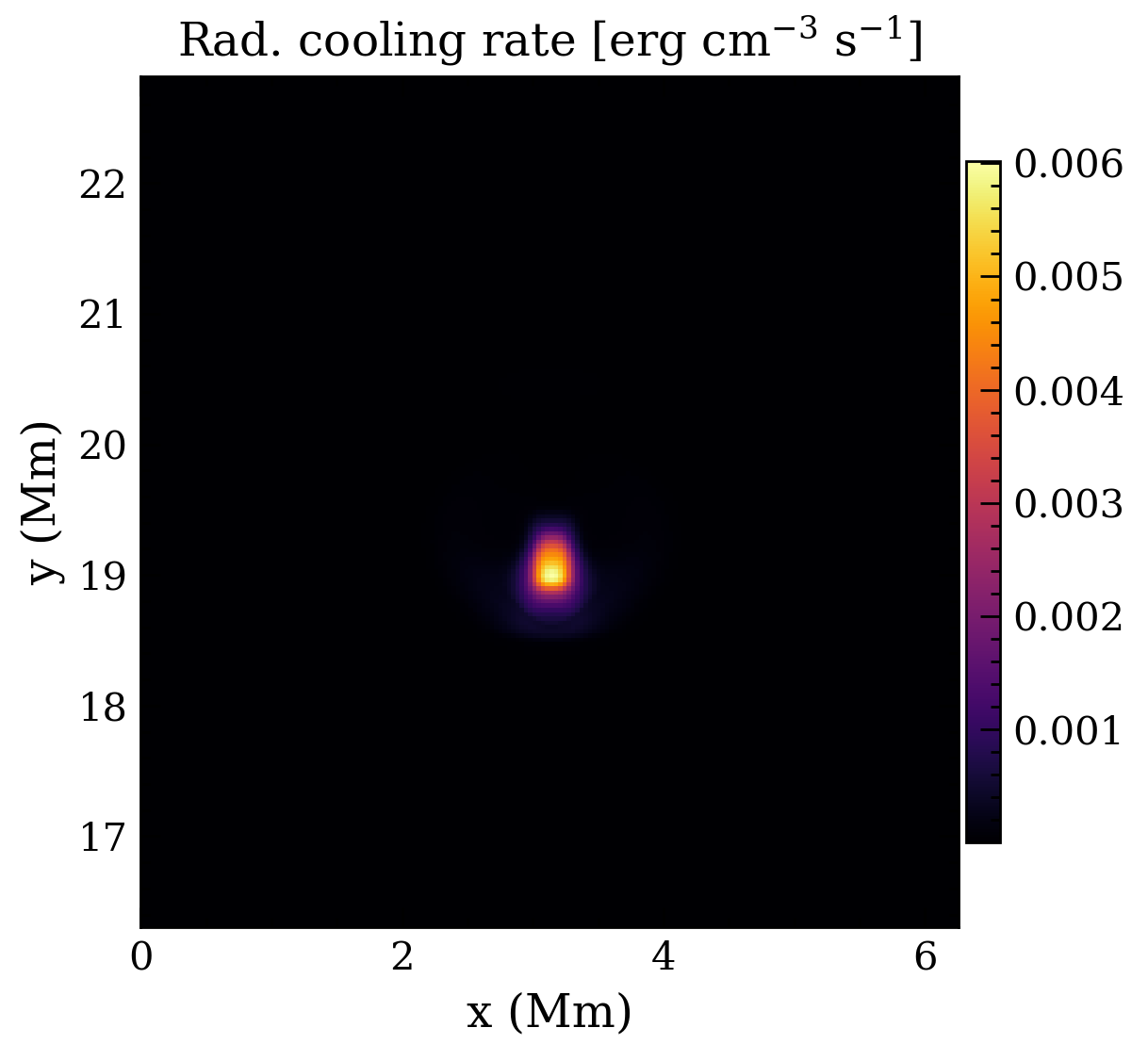}
    \caption{Spatial distribution of the cooling rate due to optically thin radiation at $t=13.38$ min. It is evident that the location of the maximum radiative loss is co-spatial with the cool-condensation site as shown in the bottom row in Figure~\ref{fig:MHD}.}
    \label{fig:RC_prom}
\end{figure}

\begin{figure}[hbt!]
    \centering
    \includegraphics[width=1\linewidth]{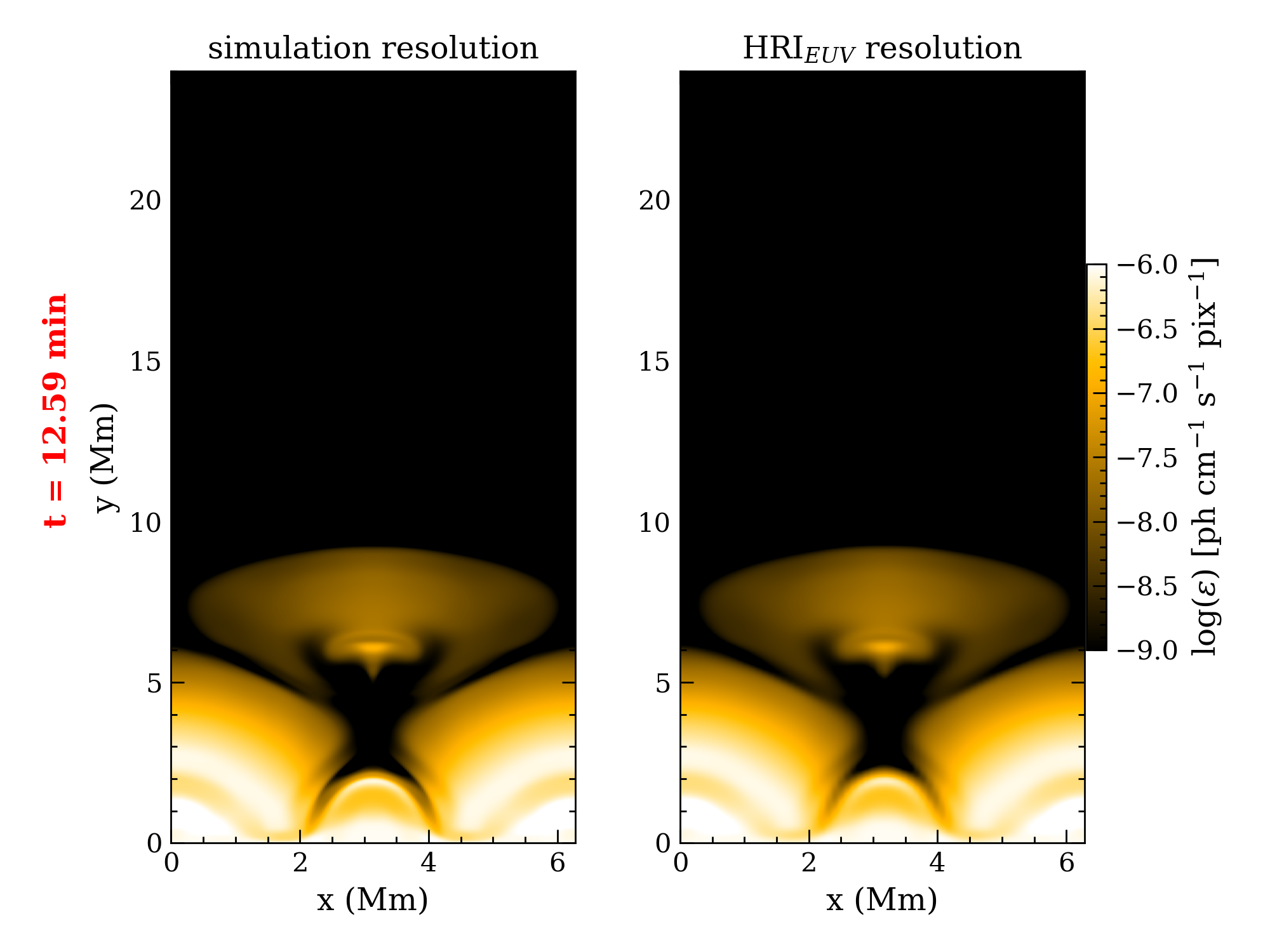}
    \includegraphics[width=1\linewidth]{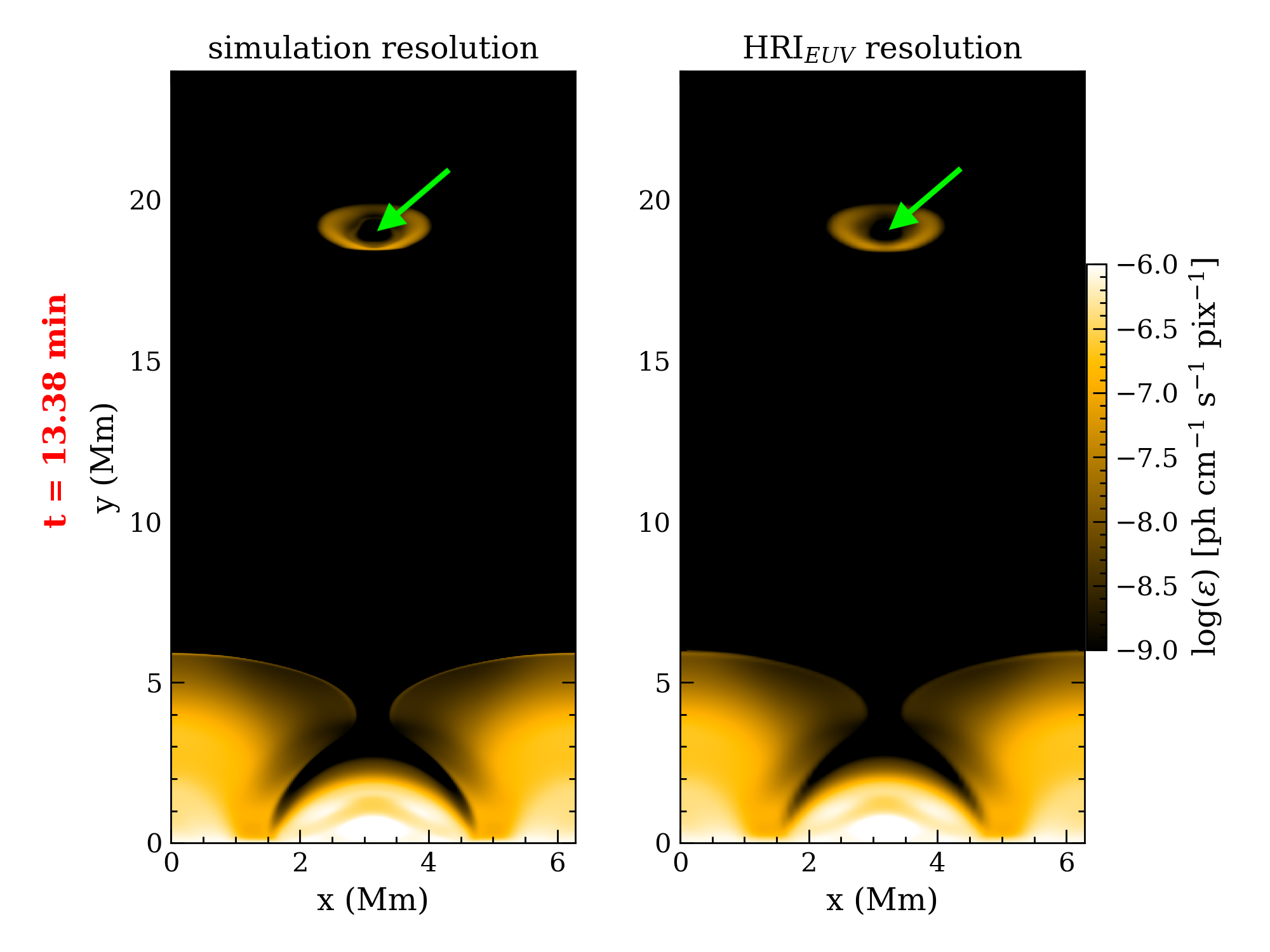}
    \caption{Synthetic emissivity maps in the SolO / HRI$_\mathrm{EUV}$ 174~\AA\ channel. The top row shows the stage when the flux rope is about to erupt ($t=12.59$ min), while the bottom row corresponds to the post-eruption stage ($t=13.38$ min). For each row, the left and right panels display the synthetic maps at the simulation resolution and at a degraded resolution comparable with that of HRI$_\mathrm{EUV}$, respectively. The dark regions inside the flux ropes as marked by the green arrows at the bottom row are the region where the cool ($\approx 40$~kK) plasma is trapped. An animation for the time span between $t=12.17$ to 13.95~min is available online.}
    \label{fig:HRI}
\end{figure}

\begin{figure}[hbt!]
    \centering
    \includegraphics[width=1\linewidth]{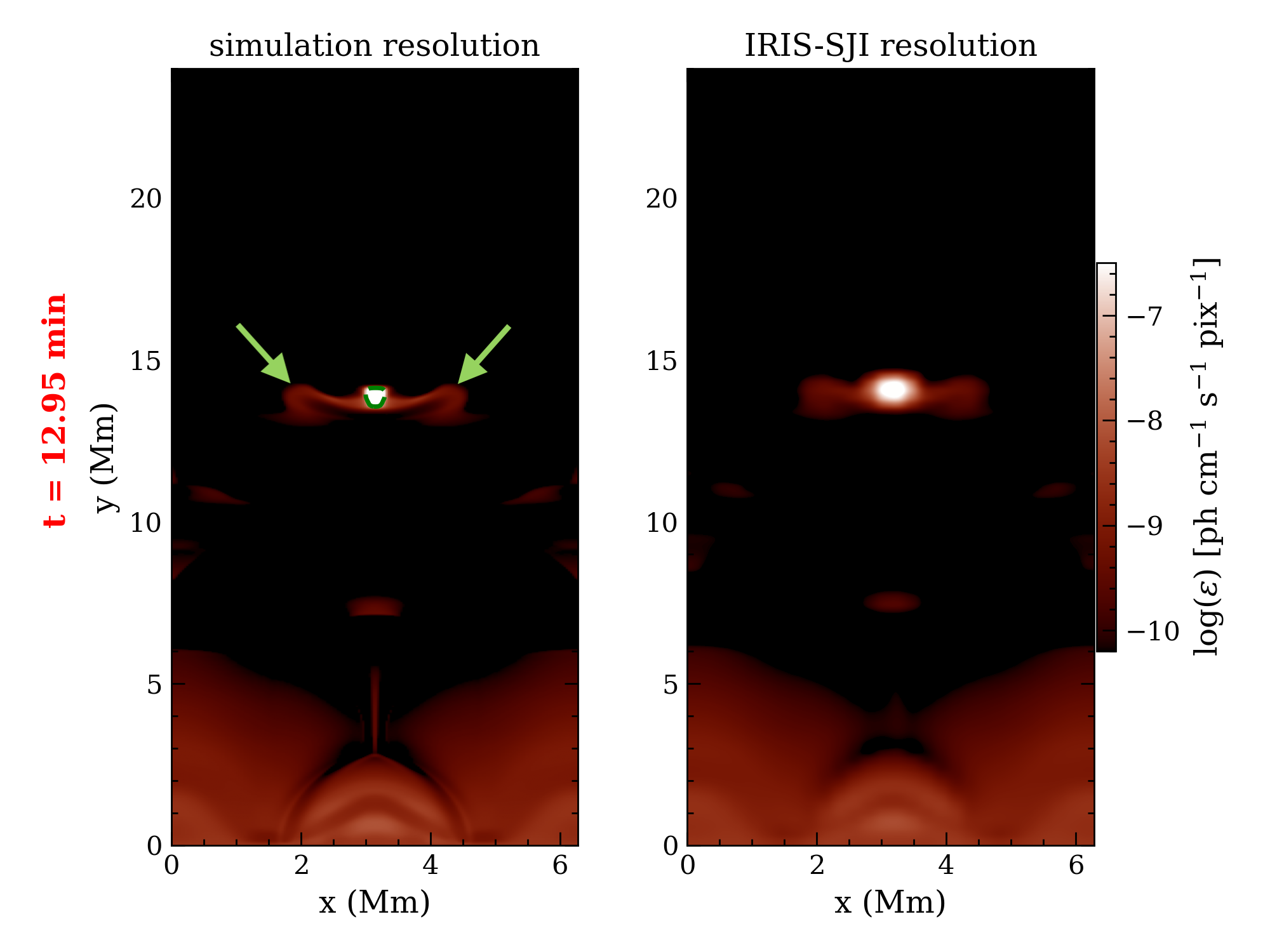}
    \includegraphics[width=1\linewidth]{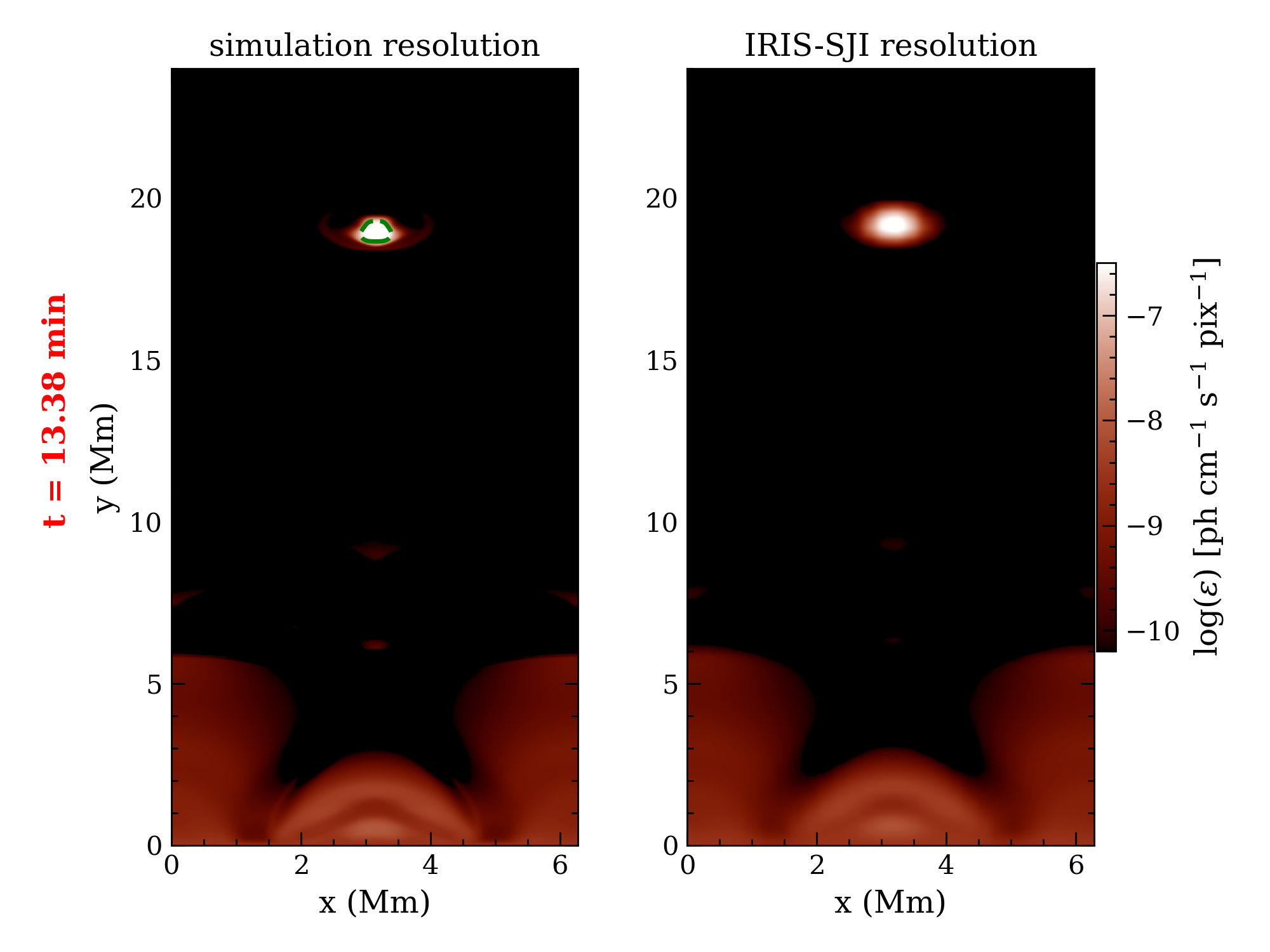}
    \includegraphics[width=0.8\linewidth]{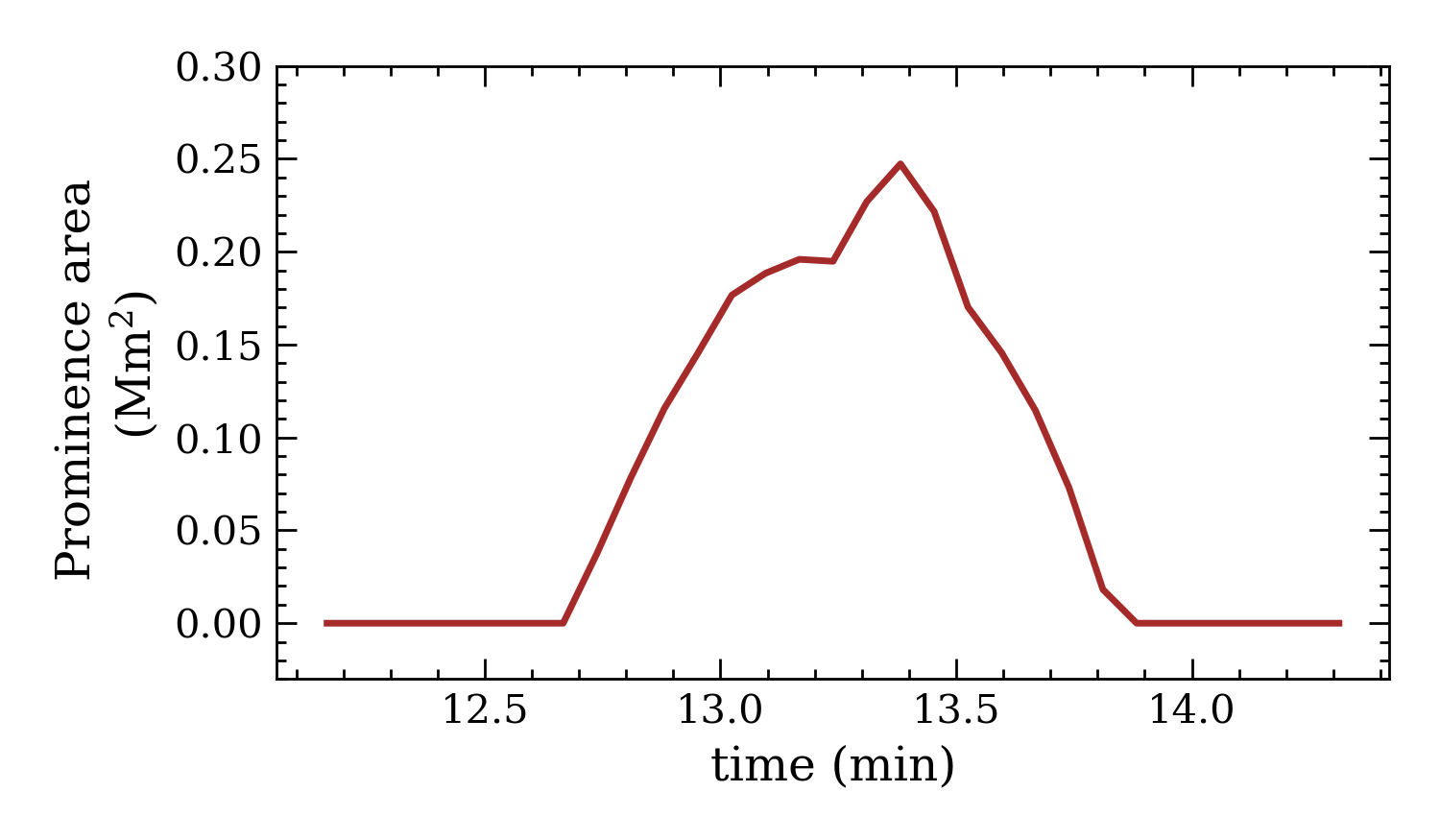}
    \caption{Top and middle rows: Synthetic emissivity map in the IRIS/SJI~1400~\AA\, passband at $t=12.95$~min and $t=13.38$~min, respectively. The left and right panels show the synthetic maps at the simulation resolution and at a degraded resolution compatible with the IRIS/SJI, respectively. An animation for the time span between $t=12.17$ to 13.95~min is available online. The green arrows marked in the top-left panel highlights the wing-like structure inferring the presence of a dip in the magnetic flux rope. Bottom row: Temporal evolution of the prominence area obtained from synthetic IRIS/SJI~1400~\AA\, imaging for a emissivity threshold ($\epsilon_{th}=9.51 \times 10^{-7}$ ph~cm$^{-1}$~s$^{-1}$~pix$^{-1}$) that correspond to 90\% of the emissivity peak value at $t=13.38$~min. The green contours represent this threshold value at the top-left and middle-left panels.}
    \label{fig:IRIS}
\end{figure}

\begin{figure}[hbt!]
    \centering
    \includegraphics[width=0.49\linewidth]{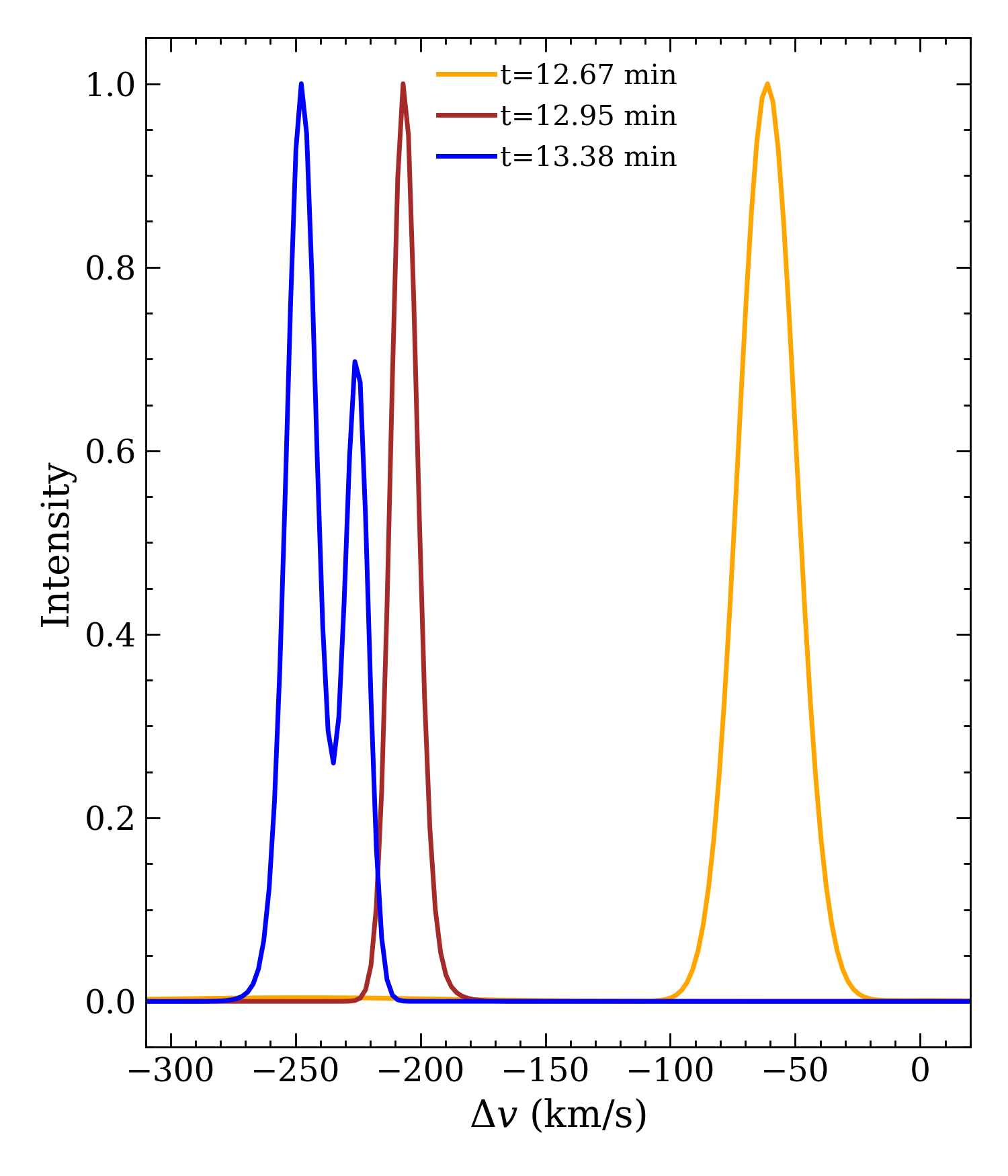}
    \includegraphics[width=0.49\linewidth]{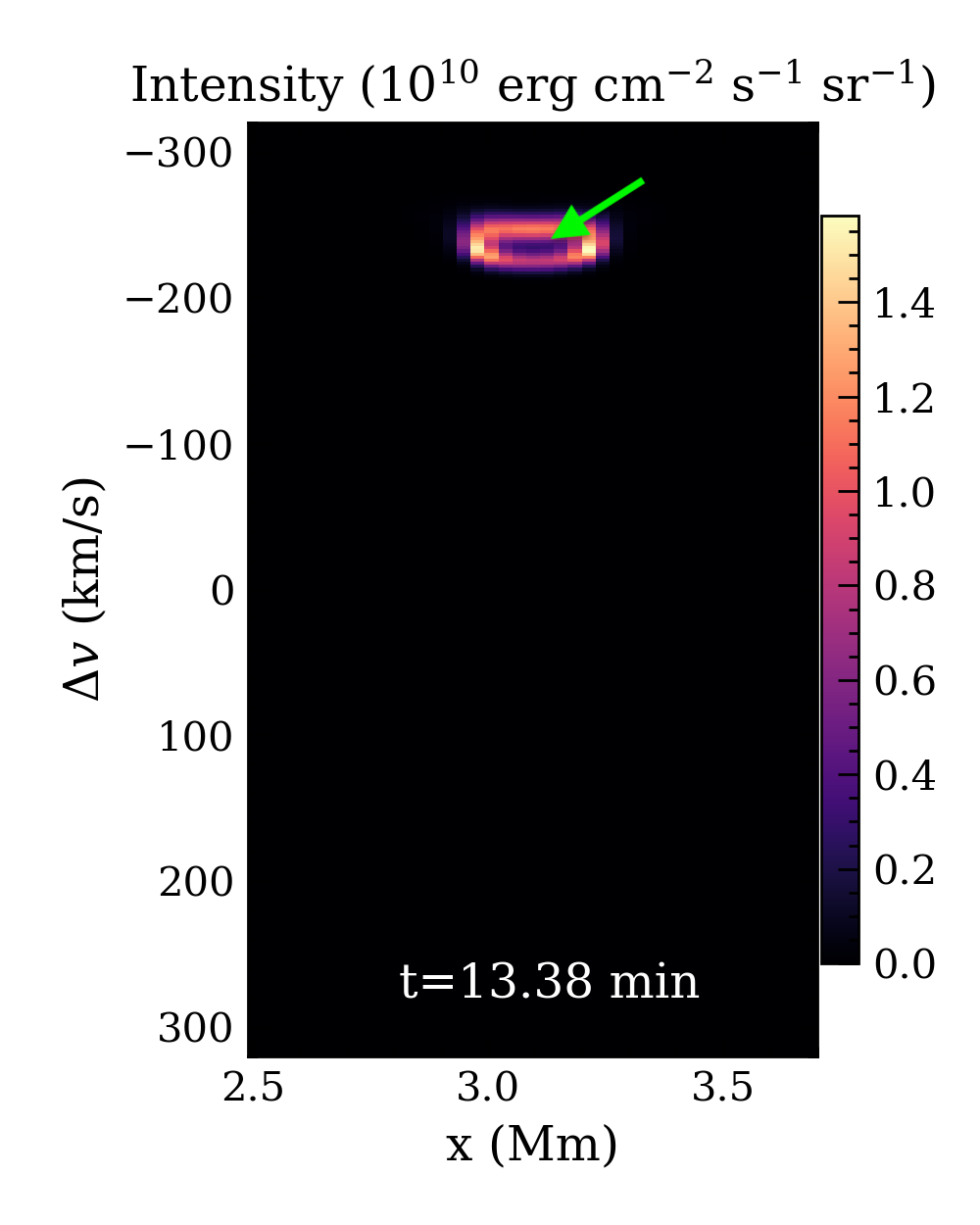}
    
    \includegraphics[width=0.48\linewidth]{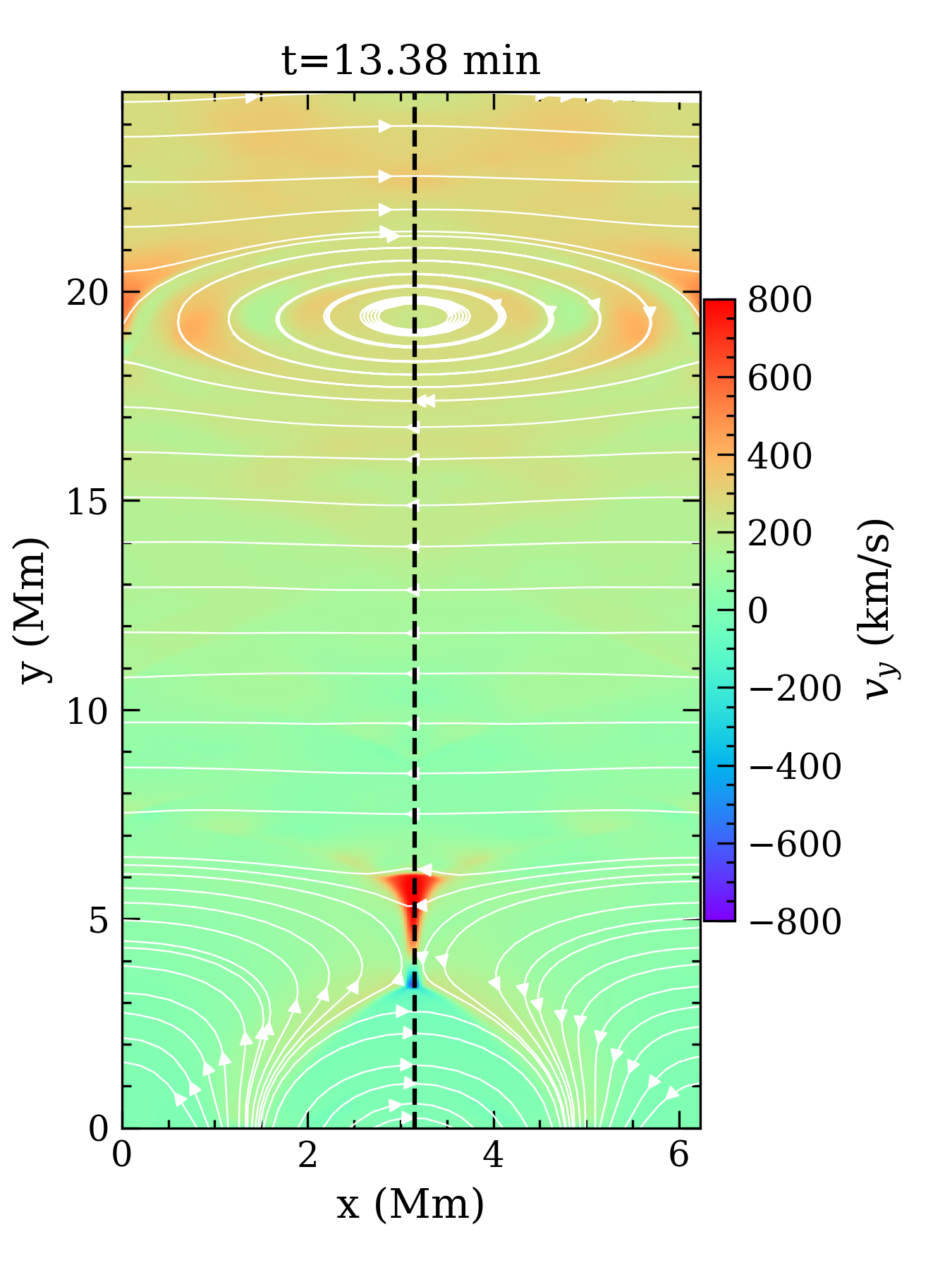}
    \includegraphics[width=0.51\linewidth]{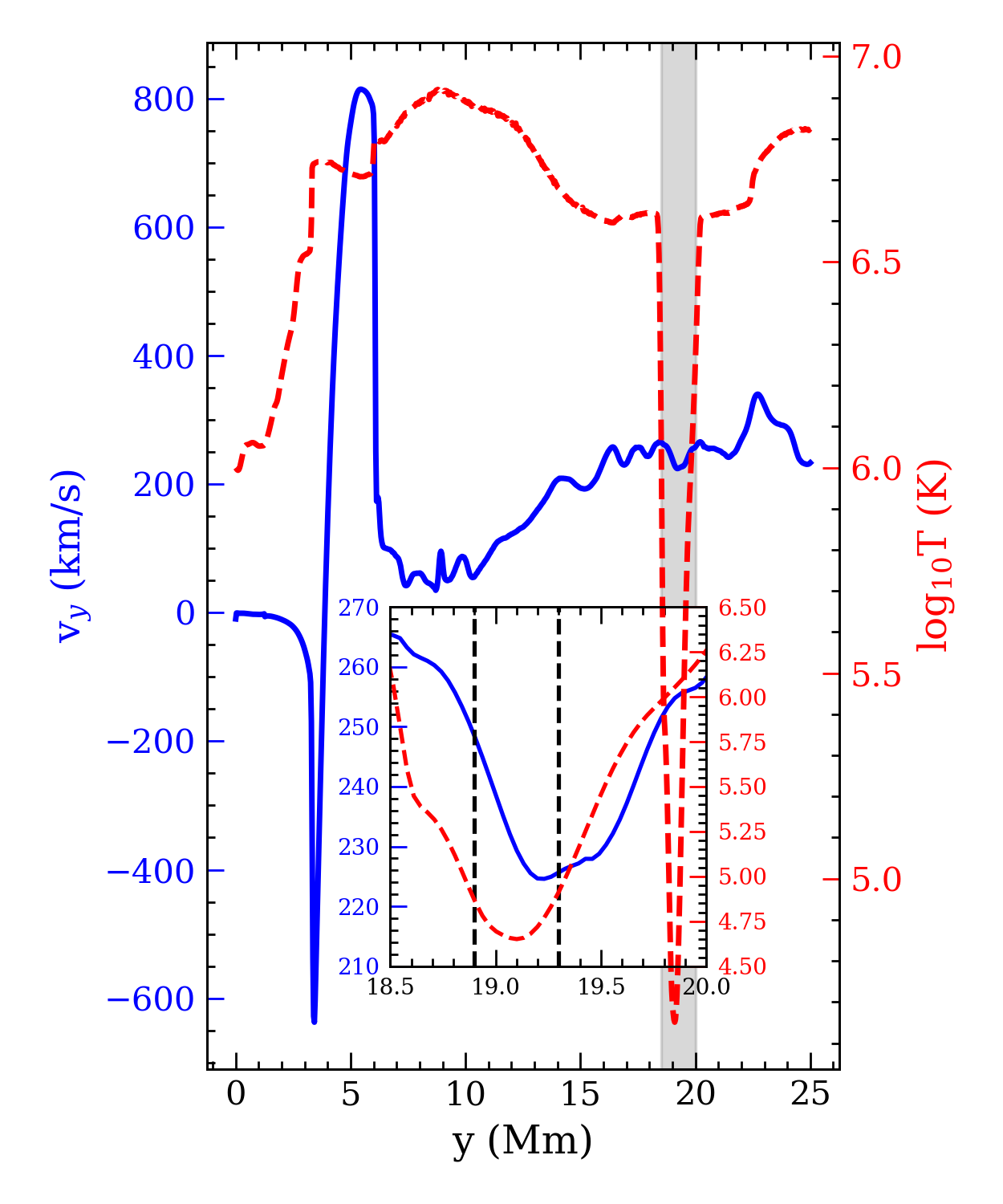}
    \caption{Top row: Synthetic spectra for Si~IV 1402.77~\AA\ for line of sight along the vertical direction ($y$-direction), where the negative and positive Doppler velocities represent the upward ($+y$ direction) and downward ($-y$ direction) motion of the plasma. The left panel shows the line spectra at $x=3.14$~Mm at three different times as shown in the legend, where the intensities at each times are normalized to unity. The presence of two predominant LOS velocity components in the spectral profile at $t=13.38$~min (blue curve) is evident, and also showcased by a green arrow in the spectral map at the right panel. Bottom row: The left panel shows the spatial distribution of the vertical velocity component ($v_y$) at $t=13.38$~min, where magnetic field lines (projected in the $x-y$ plane) are shown in white. The right panel presents the $v_y$ and temperature ($T$) distribution along the vertical cut at $x=3.14$~Mm, indicated by the black dashed line in the left panel. The inset plot highlights the region between $y=18.5-20$~Mm (shaded in gray), where the prominence material is located. The black vertical dashed lines showcases the heights at $y=18.9$ and 19.3~Mm that correspond to the temperature of 80~kK (peak formation temperature of Si~IV 1402.77~\AA\ line) and LOS velocities of $\approx 250$ and 220~km s$^{-1}$, respectively. The inset plot highlights the zoomed-in view of the gray region between $y=18.5$~Mm to 20~Mm, where the two vertical black dashed lines are the markers at $y=18.9$~Mm and 19.3~Mm.}
    \label{fig:prominence-spectra}
\end{figure}

\section{Overview of the underlying MHD model}\label{sec:simulation}
In the numerical simulation by S24 that serves as a basis for the synthesis carried out in this work, we investigate the eruption of a flux rope, and post-flare coronal rain using a 2.5D resistive-MHD simulation with \texttt{MPI-AMRVAC} \citep{2012JCoPh.231..718K, 2014ApJS..214....4P, 2018ApJS..234...30X, keppens2021, keppens2023}. The simulation domain spanned a horizontal range of $x=0-2\pi$ Mm, and a vertical range of $y=0-25$ Mm, with minimum spatial resolution of 32.6~km in either directions. The choice of the length scales in our model is motivated by the aim of reproducing small-scale MFR eruptions, which can be associated with mini-filament eruptions having widths of $\lesssim 2.2$~Mm \citep{Wang-mini-filament:2000, Li2023_mini-prominence}. However, the 2.5D geometry of our model showcases a limb-view perspective to address mini-prominence, similar to how prominences are observed at the solar limb. The initial magnetic field configuration adopted in this model is non-force-free, and replicates sheared magnetic arcades which have a guide field component along the invariant direction ($z-$axis). This approach was adopted to avoid the computational time required to develop an unstable magnetic configuration from the initial mechanical equilibrium state, which is consistent with methods used in MHD simulations initiated with non-force-free field setups, as reported by \cite{2016:sanjay, Nayak:2024} and the references therein. The initial shear angle of the projected field lines at the $y = 0$ plane with the $+x$ axis is $72.5^\circ$, and uniform throughout the simulation domain. The model was assumed to be isothermal at 1~MK at the initial stage to replicate a typical coronal medium, whereas, the vertical stratification of the plasma density and pressure maintained a hydrostatic equilibrium with solar gravity. The model incorporates Spitzer-type thermal conduction, which is strictly field-aligned, with a conductivity given by ${\bf \kappa}_{||} = 10^{-6} T^{5/2}$ erg cm$^{-1}$ s$^{-1}$ K$^{-1}$. However, there is no thermal conduction at the initial state due to the isothermal condition. Additionally, a steady background heating is included to precisely balance the initial optically-thin radiative losses, where the radiative cooling function is used from the model based on \cite{1972ARA&A..10..375D} and \cite{2008ApJ...689..585C}. The readers are referred to S24 for a detailed description of the governing equations, numerical schemes and boundary conditions used in the simulation. However, it is worthy to note that the lower boundary follows a line-tied condition, where the velocity components were set antisymmetrically, the magnetic field components were extrapolated, and the pressure and density values were set as their local initial values to fill the ghost cells.

\section{Flux rope and mini-prominence} \label{sec:MFR-prominence}
We discuss the formation and eruption of a flux rope-trapped mini-prominence based on the simulation in S24, and report their synthetic diagnostics in the following. 

\subsection{Formation and eruption}\label{sec:MFR-prominence-MHD}
Due to the initially non–force-free magnetic field configuration and the associated mechanical imbalance, the system was driven to evolve away from its initial state. By approximately $t=5.44$~min, the system reached a mechanical and thermal quasi-equilibrium, which persisted until about $t=7$~min, as explained in detail in S24. During this quasi-equilibrium phase, the guide-field component ($B_z$) contributes to the magnetic shear flux, resulting in the stretching of the upper portion ($y \gtrsim 13$~Mm) of the central arcade, as reported in \cite{Sen:2025b}. Consequently, field lines located in close proximity around $x=3.14$~Mm form a vertical current sheet (CS) region above the polarity inversion line (PIL), thereby enabling reconnection in a tether-cutting fashion \citep{Antiochos:1999, Moore:2001}. This reconnection process detaches the flux rope from the underlying arcades and triggers the first eruption at approximately $t=7.51$~min. As a result of the reconnection, the post-reconnection loops contract downward, carrying the shear flux with them and transferring magnetic flux to the underlying arcades. Subsequently, interactions between the straddling flux ropes at the side boundaries and the central arcade progressively facilitate the formation of another flux rope, and leading to a second eruption from the central arcade region at around $t=12.59$~min, in a manner similar to the first eruption \citep{Sen:2025b}.

To appreciate the location of the flux rope at different heights at different times, we showcase the distribution of the plasma density ($\rho$) and temperature ($T$) in the vicinity of the flux rope in Figure~\ref{fig:MHD}. From top to bottom panels, it represent the three stages, respectively: shortly after its formation (at $t=12.59$~min), and later stages during the eruption at $t=12.95$~min, and $t=13.38$~min. At $t=12.59$~min, a localized density enhancement is evident within the flux rope, as shown in the top-left panel. The central part of the magnetic arcades (for $y\lesssim 10$~Mm) are highly sheared (with a maximum value of $90^\circ$) and nearly parallel to the polarity inversion line that is present at $x=3.14$~Mm at the bottom boundary, as shown in \cite{Sen:2025b}. This magnetic structure can support plasma materials without falling due to gravity, and contributing to the formation of filament channel, also reported in \cite{Kinizhink:2015, zhao2017, Kinizhink:2017}. During the eruption stages at $t=12.95$~min and $t=13.38$~min, the material is scooped up by the flux rope and progressively accumulates within a more confined region due to presence of the dip in the magnetic field structure, as illustrated in the middle-left and bottom-left panels, respectively. At $t=13.38$~min, the maximum density of the localized plasma reaches a value $\gtrsim10$ times higher than the surrounding background medium (see bottom-left panel of Figure~\ref{fig:MHD}), and is associated with temperatures of approximately 40~kK, as shown in the bottom-right panel. This cool plasma structure does not get sufficient time to undergo further cooling and may also experience adiabatic expansion as it erupts together with the flux rope. This cool-condensation corresponds with the observed range of prominence plasma, which is typically $\sim 10^4$~K, and 10 to 100 times denser than the surrounding corona \citep[and references therein]{Mackay:2010, Parenti:2014}. 

To investigate the interplay between heating and cooling rate at the cool site as shown in the bottom-right panel in Figure~\ref{fig:MHD}, we estimate the energy loss rate due to optically thin radiation. The spatial variation of the radiative energy loss rate is shown in Figure~\ref{fig:RC_prom}. It shows that the cooling rate has a maximum value of $\approx 6 \times10^{-3}$ erg cm$^{-3}$ s$^{-1}$ which is localized at around $y=19$~Mm and $x=3.14$~Mm, and co-spatial with the cool-condensed site. This cooling rate is much more dominant than the steady background heating rate, which has a value of $2.9 \times 10^{-3}$ erg cm$^{-3}$ s$^{-1}$ at the same location. This implies that the radiative cooling becomes more effective than the net heating, which leads to the formation of the cool-site.    

\subsection{Synthetic observations in EUV and UV}\label{sec:syn_obs_euv}
We use an optically-thin approximation to obtain synthetic observations in the extreme ultraviolet (EUV) and ultraviolet (UV) channels, compatible with the Solar Orbiter (SolO) / High Resolution Imager \citep[HRI$_{EUV}$;][]{SolO-EUI:2020} and Interface Region Imaging Spectrograph \citep[IRIS;][]{IRIS:2014}, respectively. The detailed methodology for synthetic imaging and line spectra for optically thin lines are described in \cite{Sen:2026}. 

\subsubsection{Imaging}\label{sec:imaging}
We use the response function for SolO/HRI$_{EUV}$, which corresponds to the Fe IX~174~\AA\ line for a temperature range between $10^5$ to $10^8$~K with a constant density of $10^9$~cm$^{-3}$, and assuming coronal abundance from \cite{DelZanna:2023}. This response function is obtained through private communication with the Solar Orbiter team, which is also used in other reports \citep[e.g.,][]{Faerder_solo:2024, Daniel_solo:2025}, that has a default unit in DN~cm$^5$~s$^{-1}$~pix$^{-1}$. We convert the electron number density ($n_e$) from the plasma density ($\rho$) assuming a fully ionized plasma with $10:1$ abundance of H and He as $n_e = \frac{\rho}{1.4 m_p}$, where, $m_p$ denotes the proton mass \citep{Xia:2012}. To convert the response function in photon counts, we multiply by a factor of 1/7 to obtain into the ph~cm$^5$~s$^{-1}$~pix$^{-1}$ unit \citep{solo-EUI_calibration:2026}. To estimate the response function for the IRIS/Slit Jaw Imager (IRIS/SJI) 1400~\AA\ passband, we follow the methodology as described in \citet{Antolin:2026} (we refer the reader to this paper for a more detailed description). First, the contribution function, $G(n, T)$ is obtained for a passband in the range between 1301.5 and 1444.5~\AA\ for Si~IV, in a temperature ($T$) range between $10^4$ to $10^{7.5}$~K, and number density ($n$) between $10^7-10^{13}$~cm$^{-3}$ using the \texttt{iris\_get\_response.pro} routine from the \texttt{SSWIDL} package. Here the effect of charge transfer is included using the CHIANTI~v.11 \citep{Dere_1997AAS..125..149D, CHIANTI11:2024}, and the photospheric abundance from \cite{Asplund:2021} is used. Next, the effective area and the ratio between proton to electron density are included to obtain the contribution function ($G_{ph}(n, T)$) in the unit of ph~cm$^5$~s$^{-1}$~sr$^{-1}$. Finally, the response function ($R_{ph}(n, T)$) in the unit of ph~cm$^5$~s$^{-1}$~pix$^{-1}$ is obtained by
\begin{align}\label{eq:IRIS_response}
    R_{ph}(n,T) = G_{ph}(n,T) \times  \bigg(\frac{\pi \times 180}{3600}\bigg)^2 \times \theta^2_{ps},
\end{align}
where, $\theta_{ps}=0''.166$ is the plate-scale for the IRIS/SJI.

To showcase the imaging signature of the flux rope during its formation and the eruption stages, we synthesize the emissivity maps in the EUV channel compatible to the SolO/HRI$_{EUV}$ at two different times at $t=12.59$ and 13.38 mins, as presented in the top and bottom rows of Figure~\ref{fig:HRI}, respectively. The left column in each rows represent the synthetic maps at the simulation resolution, whereas, the right column at each row represent the corresponding synthetic counterparts degraded with a pixel resolution of 110 km, consistent to the SolO/HRI$_{EUV}$ pixel resolution at its perihelion distance \citep{Faerder_solo:2024}. The resolution degradation is performed using a Gaussian convolution. The brightening between $y=5$ to 9 Mm that appears in the emissivity map at the top-left panel in Figure~\ref{fig:HRI} shows the signature of a flux rope which is just about to erupt. This feature is well discernible in the corresponding map at the instrumental resolution of SolO/HRI$_{EUV}$, as shown in the top-right panel in the same figure. At a later time at $t=13.38$~min, the rim-like brightening is the signature of the erupting flux rope, which is evident at around $y=19$~Mm, as shown in the bottom row of Figure~\ref{fig:HRI}.  

To capture the signature of the erupting cool material that is trapped within the flux rope, we synthesize the emissivity map in IRIS/SJI 1400~\AA\ passband, which is dominated by the Si IV~1402.77~\AA\ emission, and has a peak formation of $\log T = 4.8$~K, as shown in the top and middle rows of Figure~\ref{fig:IRIS}. The left and the right panels represent the synthetic maps at the simulation resolution and the resolution of the instrument ($0''.33$ per pixel), respectively. The localized cool material at around $y=14$~Mm is trapped within the flux rope at $t=12.95$~min. The wing-like structure in the top-left panel, marked by the green arrows beneath the brightest region, indicates the presence of the magnetic dip within the flux rope where the cool plasma accumulates. This cool region has a temperature close to the peak formation temperature of IRIS/SJI 1400~\AA\ (80~kK), with a minimum value of $\approx 72$~kK, while remaining embedded within plasma at temperatures of $\gtrsim$1~MK. As shown in the middle row of Figure~\ref{fig:MHD}, the plasma density in this cool region is approximately 10 times greater than that of its surroundings. Consequently, the condensed material within the flux rope attains an enhanced emissivity ($\epsilon \sim n^2 R_{ph}(n, T)$) relative to the ambient plasma, producing the bright feature seen in the emissivity map. In contrast, the surrounding hotter plasma ($\approx 1$~MK) appears as fainter wing-like features, as indicated by the green arrows. This is attributable to the weaker emission resulting from its lower density structure, as shown in the middle-left panel of Figure~\ref{fig:MHD}, additionally with the substantially weaker emission from Fe XII within the passband (see \cite{Antolin:2026}). At a later stage, at $t=13.38$~min, the flux rope-trapped cool material reaches a height of $y\approx19$~Mm during the eruption. At this time, it exhibits temperatures of $\approx 40$~kK and a plasma density approximately 10 times higher than the background plasma (see the bottom row of Figure~\ref{fig:MHD}). As a result, the cool condensed material produces the bright structure located around $y=19$~Mm, which is surrounded by a fainter background, as shown in the middle row of Figure~\ref{fig:IRIS}. These bright regions are also co-spatial with the dark core regions within the erupting flux rope as marked by the green arrows in the HRI$_{EUV}$ maps at the bottom row of Figure~\ref{fig:HRI}. This feature corresponds to a mini-prominence which erupts along with the flux rope. Therefore, the multi wavelength analysis of SolO/HRI$_{EUV}$ and IRIS/SJI synthetic imaging underscores the necessity, in a way, to complement each other, where the signatures of the flux ropes and mini-prominence materials are evident in the EUV and UV bands respectively. 

We estimate the temporal evolution of the area of the erupting mini-prominence from the emissivity map (with simulation resolution) of IRIS/SJI, which serves as a guide to estimate the evolution of the prominence size from observation. For that, we first select a spatial domain between $x=2$ to 4.28~Mm and $y=3$ to 25~Mm. Then, at each time frame, we identify the pixels whose emissivity exceeds a threshold value. We choose the threshold $\epsilon_{th} = 9.51 \times 10^{-7}$~ph cm$^{-1}$ s$^{-1}$ pix$^{-1}$, which corresponds to the 90\% of the peak emissivity at $t=13.38$~min (marked by the green contours in Figure~\ref{fig:IRIS}). This  corresponds to the intensity of 951~ph~s$^{-1}$~pix$^{-1}$, assuming the bright feature extends up to 10~Mm along the invariant direction. The choice of the mini-prominence size along the invariant direction is motivated from the observation of an erupting mini-filament \citep{Li2023_mini-prominence}, which has the length of 30~Mm. The total area at each time step is then computed by multiplying the number of selected pixels by the cell area of the native simulation resolution ($32.6$~km$^2$). The resulting time profiles is shown in the bottom panel of Figure~\ref{fig:IRIS}. It shows the gradual growth of the size of the mini-prominence starting at approximately $t=12.7$~min, reaching a maximum value of the area of 0.25~Mm$^2$ at $t\approx13.4$~min. Subsequently, the structure becomes progressively more spatially confined, leading to a decrease in area, and eventually exits the vertical domain at around $t=13.9$~min. An animation illustrating the evolution of the mini-prominence (associated with Figure~\ref{fig:IRIS}) is available online. It should be noted that the effects of background, dark, and readout noises are not included in the synthesis. Therefore, assuming an exposure time of 4~s, which is comparable to the simulation cadence (4.29~s), the signal-to-noise ratio ($S/N \approx \sqrt{I_{ph} \ t_{exp}}$, where, $I_{ph}$ is the intensity in photon counts, and $t_{exp}$ is the exposure time) is $\approx 60$.            

\subsubsection{Spectral line}\label{sec:spectral_line}
We perform the spectral line synthesis in Si~IV 1402.77~\AA\ line under the optically-thin approximation by taking the line of sight (LOS) integration starting from the top of the simulation domain and proceeding downwards to the bottom boundary along the $y$-direction. This corresponds to a viewing direction along the solar disk, and therefore, we call it `filament view' hereafter. We compute the contribution function for Si~IV 1402.77~\AA\ line using the \texttt{ch\_synthetic.pro} routine from CHIANTI~v.10 \citep{DelZanna:2021}, and adopting the photospheric abundances from \cite{Asplund:2021}. The contribution function is calculated over a temperature range from $10^4$ to $10^{7.5}$~K, and a density range between $10^5$ to $10^{13}$~cm$^{-3}$. While it is often assumed that the Si~IV lines are formed under optically-thin conditions, this might not always be the case, in particular during extreme energetic events \citep{Kerr:2019, Zhou_IRIS:2022}. The spectral synthesis of the Si~IV 1402.77~\AA\ line considering the effect of non-equilibrium ionization is out of the scope of the current work, however, it might be interesting to investigate whether (and how much) the inclusion of this effect would modify the result in the future studies, specifically when the characteristic time scales of heating or cooling are substantially short. We adopt a spectral sampling of 26~m\AA, consistent with the IRIS spectrograph. We supply the plasma density, temperature, and LOS velocity (i.e., $v_y$) from the simulation, and use the contribution function to perform the spectral synthesis. The conversion between the electron number density ($n_e$) and the plasma density ($\rho$) follows the same prescription as mentioned in section~\ref{sec:imaging}. The top-left panel in the Figure~\ref{fig:prominence-spectra} shows the synthetic line profiles at $x=3.14$~Mm at three different times: $t=12.67$, 12.95, and 13.38~min, showcasing a blue shift from the line core. The peak of the line profile at $t=12.67$~min represents an upward motion of the mini-filament material with a velocity of around 50~km~s$^{-1}$ at the early stage of the eruption. While at the advanced eruption stage when the flux rope accelerates at $t=12.95$~min (brown curve), the velocity increases to $\approx 200$~km~s$^{-1}$. At a later time $t=13.38$~min, the line profile (blue curve) exhibits a bimodal distribution with $\Delta v \approx -220$ and -250~km~s$^{-1}$, indicating the presence of two distinct dominant velocity components of the emitting plasma along the LOS. This feature is appreciated in the corresponding spectral map at the right panel of the same figure, and marked by a green arrow. 

To illustrate the spatial distribution of the LOS velocity component at $t=13.38$~min, we present the $v_y$ map of the entire simulation domain in the bottom-left panel of Figure~\ref{fig:prominence-spectra}. The bidirectional (upward and downward) plasma flows ($\lesssim 800$~km~s$^{-1}$) are evident around a height of $y=4$~Mm, resulting from magnetic reconnection at $x=3.14$~Mm due to the interaction of the field lines of the straddling side arcades. The variation of the LOS velocity (and temperature) along the vertical cut at $x=3.14$~Mm is shown in the bottom-right panel, where the gray-shaded region corresponds to heights between $y\approx 18.5-20$~Mm (at $x=3.14$~Mm). This region exhibits velocities of $\approx 220$-270~km~s$^{-1}$ and contains the mini-filament trapped inside the erupting flux rope (see the bottom row of Figure~\ref{fig:MHD}). The intensity peaks in the Si IV 1402.77~\AA\ spectral line profile at $t=13.38$~min (blue curve at the top-left panel of Figure~\ref{fig:prominence-spectra}) arise from a complex combination of density, temperature, LOS velocity, and the contribution function. Although the LOS velocity at the mini-filament location is sufficiently lower than in the vicinity of the reconnection site (at $y\approx 4$~Mm), the dominant contribution to the spectral line intensity comes from the mini-filament region, where the temperature approaches the peak formation temperature ($80$~kK) of the Si IV 1402.77~\AA\ line, as shown by the two vertical black dashed lines at $y=18.9$ and 19.3~Mm in the inset plot at the bottom-right panel of Figure~\ref{fig:prominence-spectra}, and correspond to the LOS velocity of $\approx 250$ and 220~km~s$^{-1}$, respectively. In contrast, the plasma at other heights ($\gtrsim 1$~MK) contributes much more weakly to the intensity. Therefore, the bimodal peaks in the spectral profile represent the Doppler shifts originating from the vicinity of the mini-filament region. We also note that the line profile at $t=12.67$~min (orange curve) shown in the top-left panel is slightly more broadened along the $\Delta v$ axis, representing a higher thermal broadening than the cases at $t=12.95$ and 13.38~mins. This is expected as the temperature of the filament material during the earlier stage of eruption at $12.67$~min is hotter than the advanced stages when it becomes cooler during eruption.  

\begin{figure*}[hbt!]
    \centering
    \includegraphics[width=0.3\linewidth]{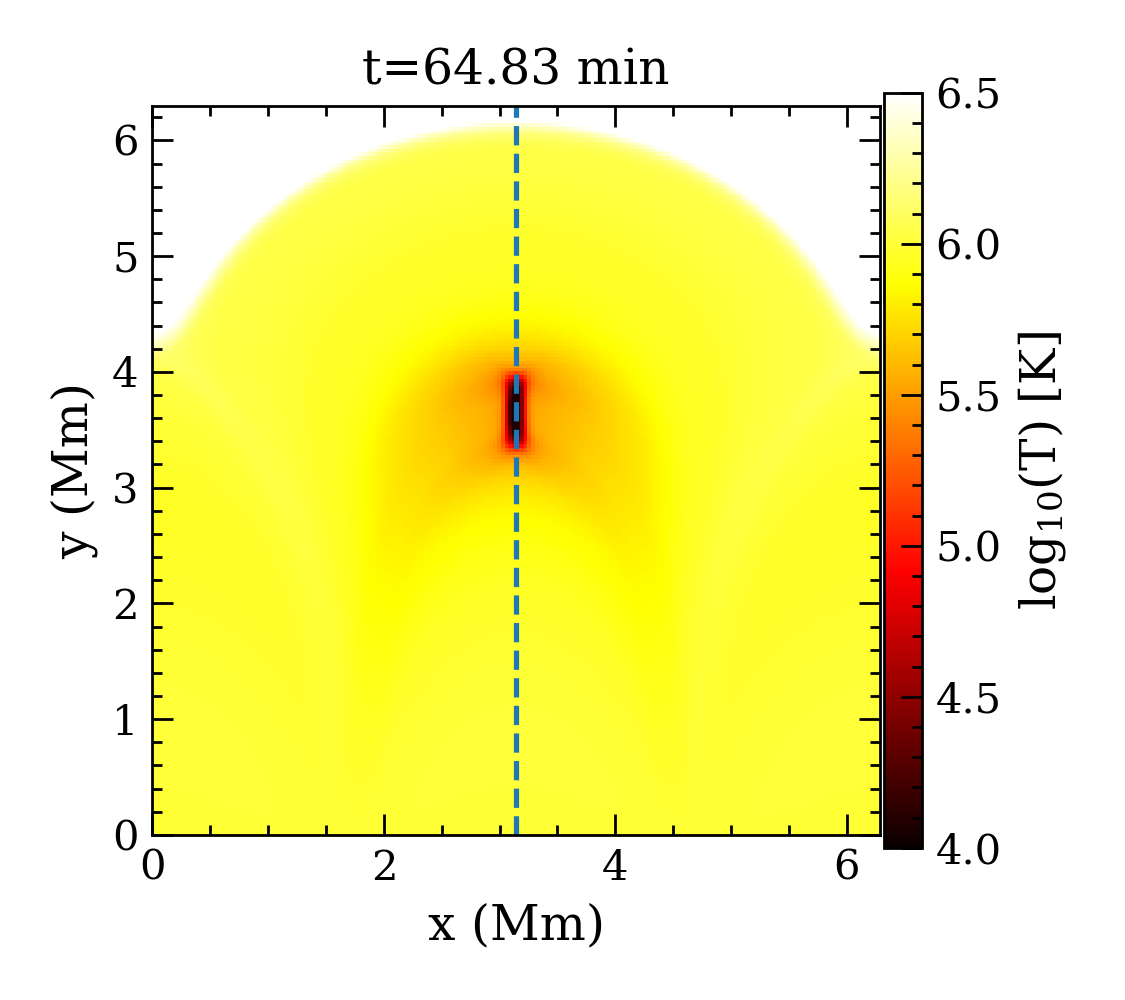}
    \includegraphics[width=0.3\linewidth]{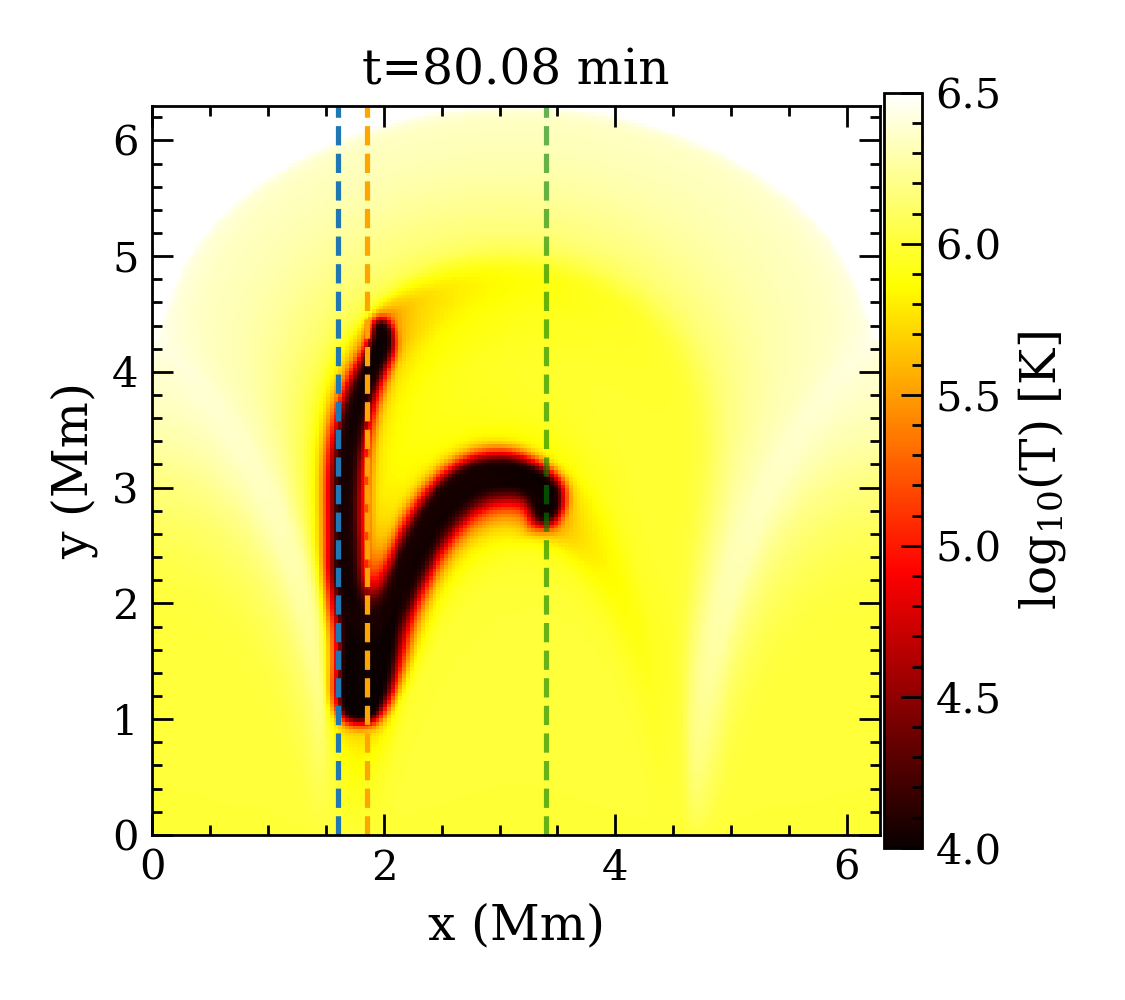}
    \includegraphics[width=0.3\linewidth]{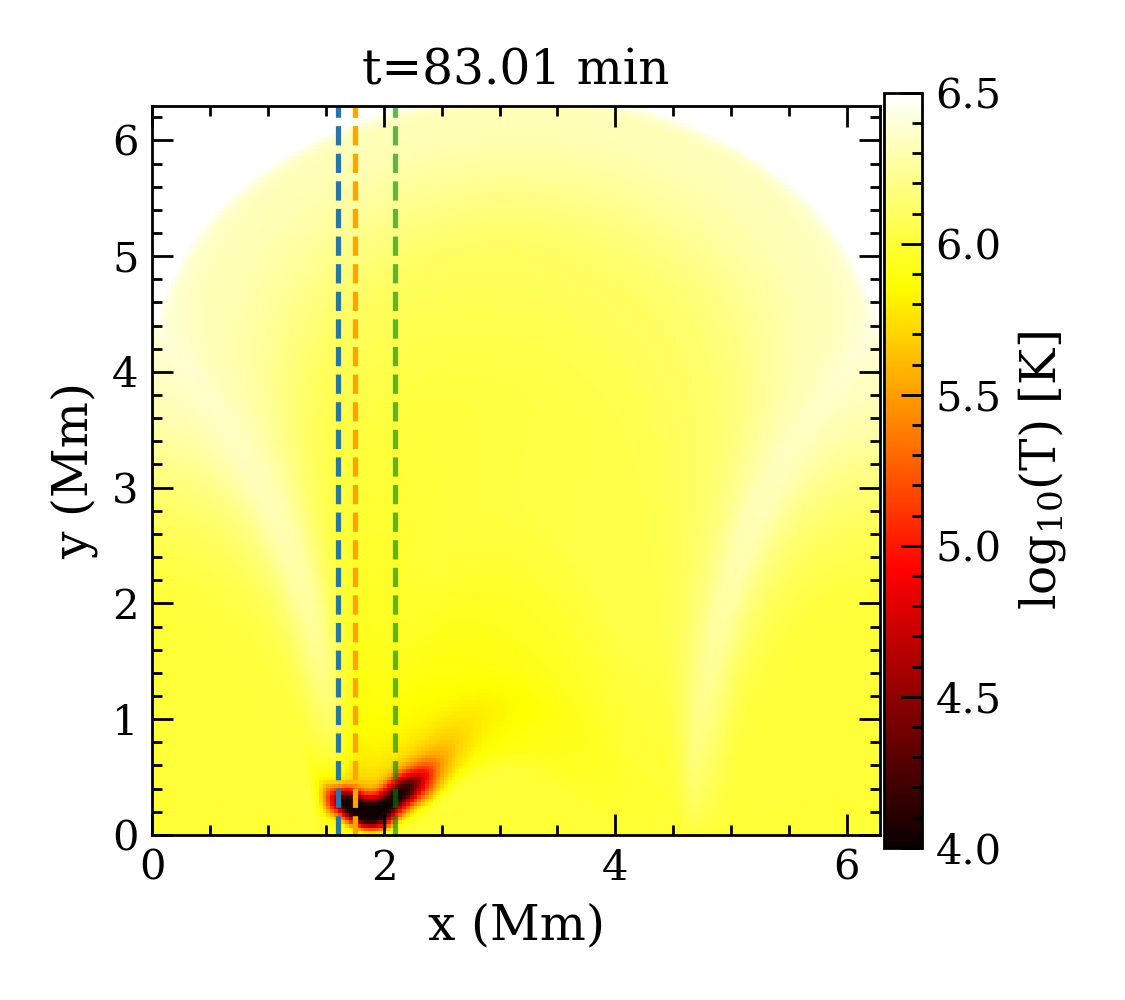}

    \includegraphics[width=0.3\linewidth]{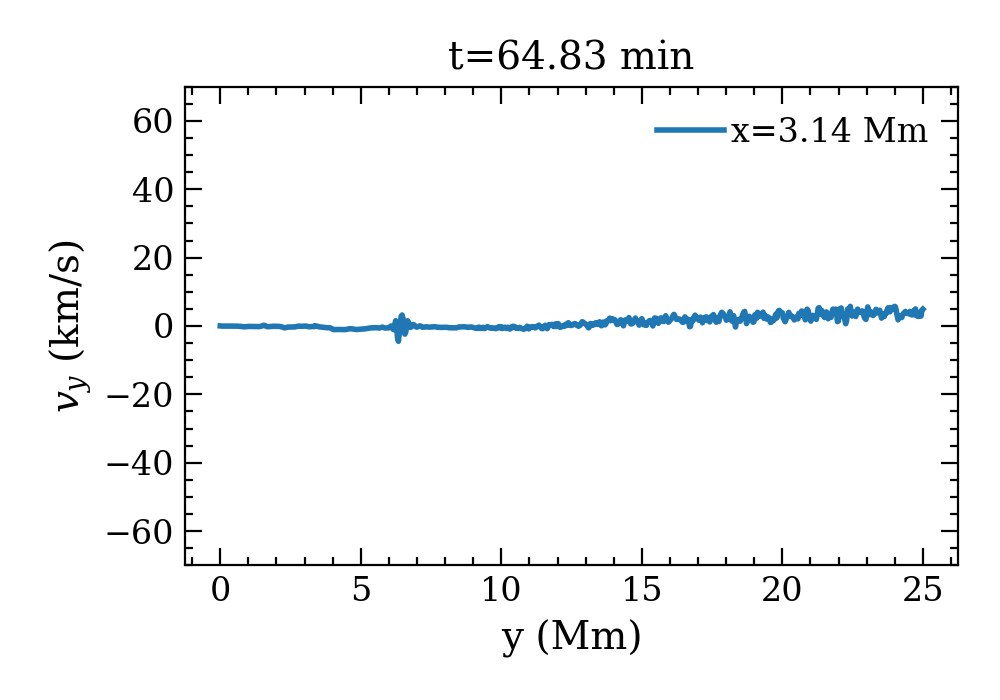}
    \includegraphics[width=0.3\linewidth]{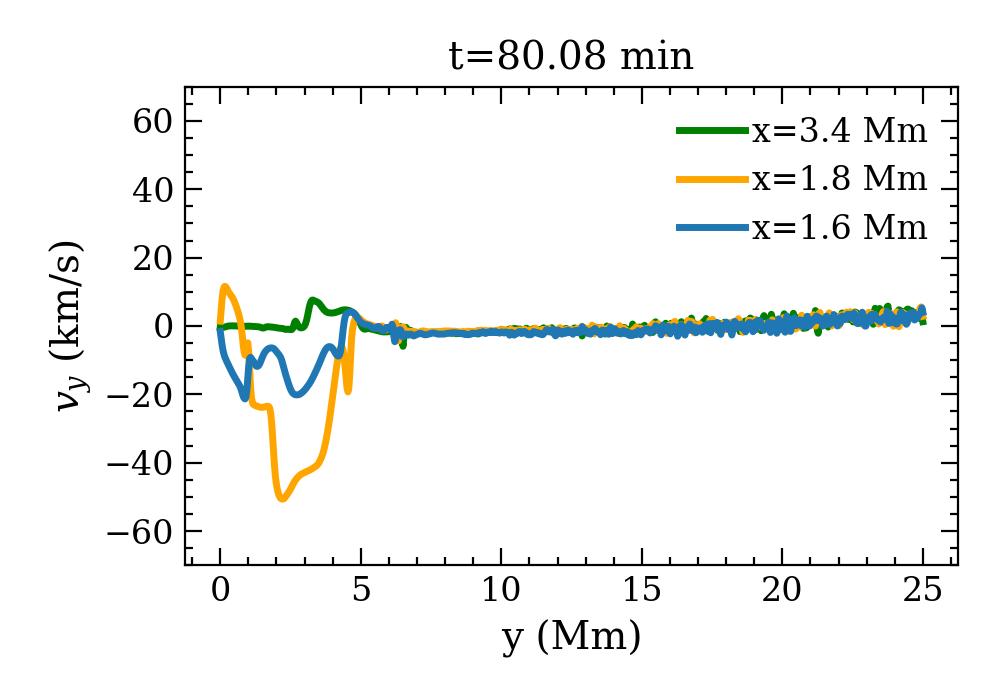}
    \includegraphics[width=0.3\linewidth]{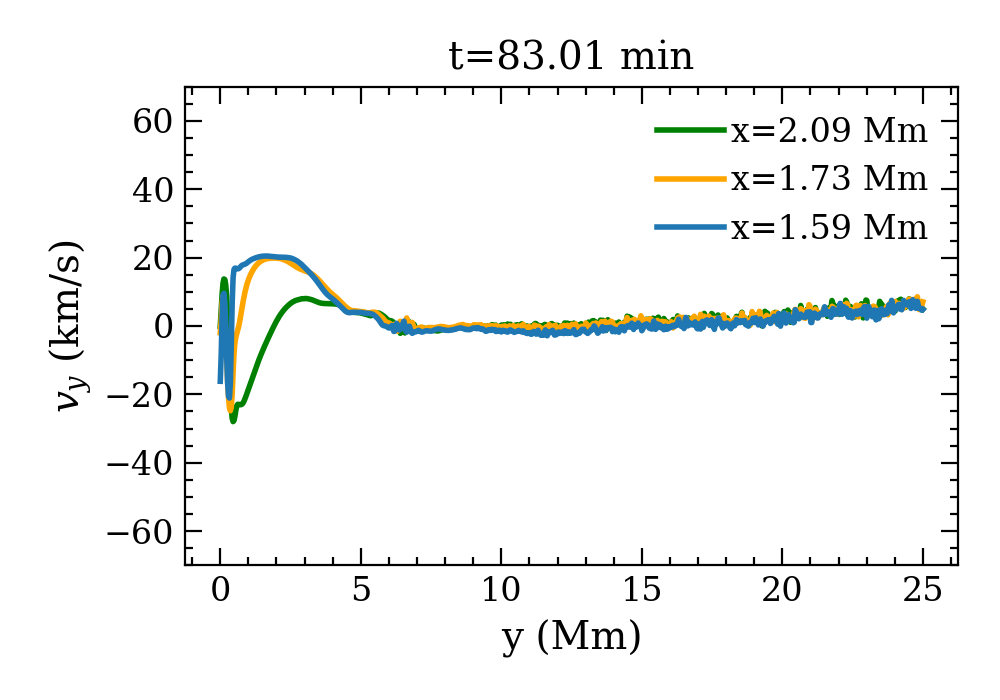}
    \caption{Top row: Temperature maps for different stages of the post flare coronal rain at three different times at $t=64.83$, 80.08, and 83.01~mins as shown from the left to right panels respectively. The vertical dashed lines at these three stages are selected to pass through different portions of the coronal rain blobs. Bottom row: $v_y$ distribution as a function of height at different $x$ locations, as indicated by the corresponding legends. The line colors are chosen to match those of the vertical dashed lines marked in the top-row maps at each time.}
    \label{fig:MHD_rain}
\end{figure*}

\begin{figure*}[hbt!]
    \centering
    \includegraphics[width=0.75\linewidth]{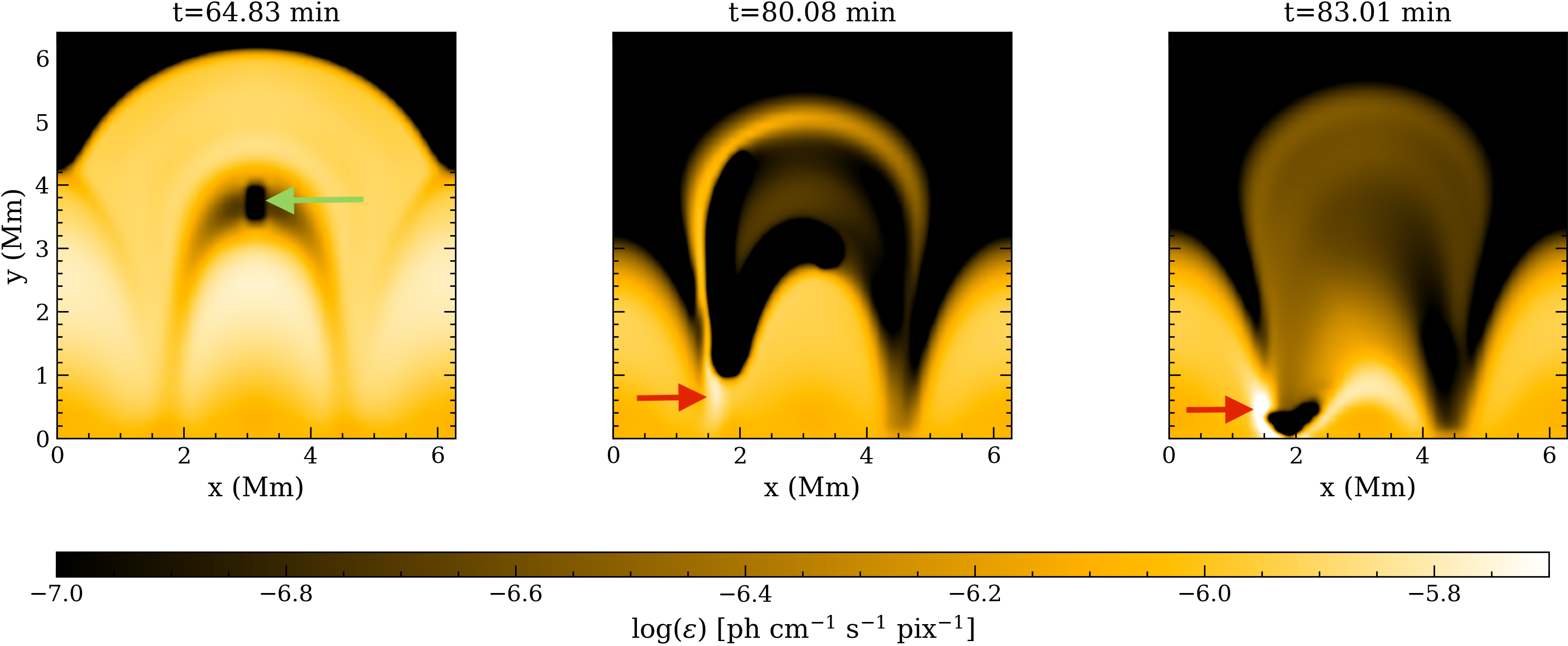}
    \includegraphics[width=0.39\linewidth]{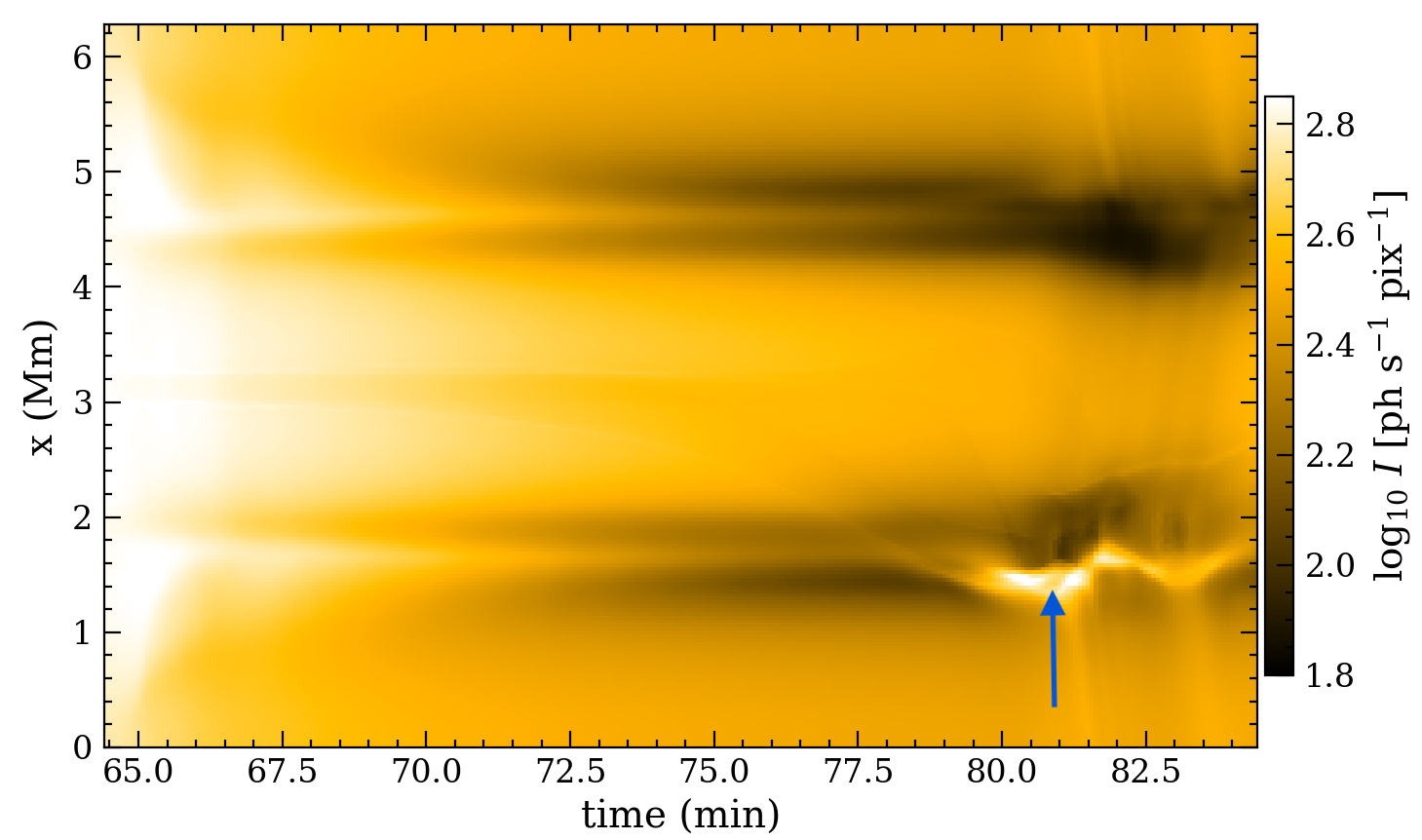}
    \includegraphics[width=0.35\linewidth]{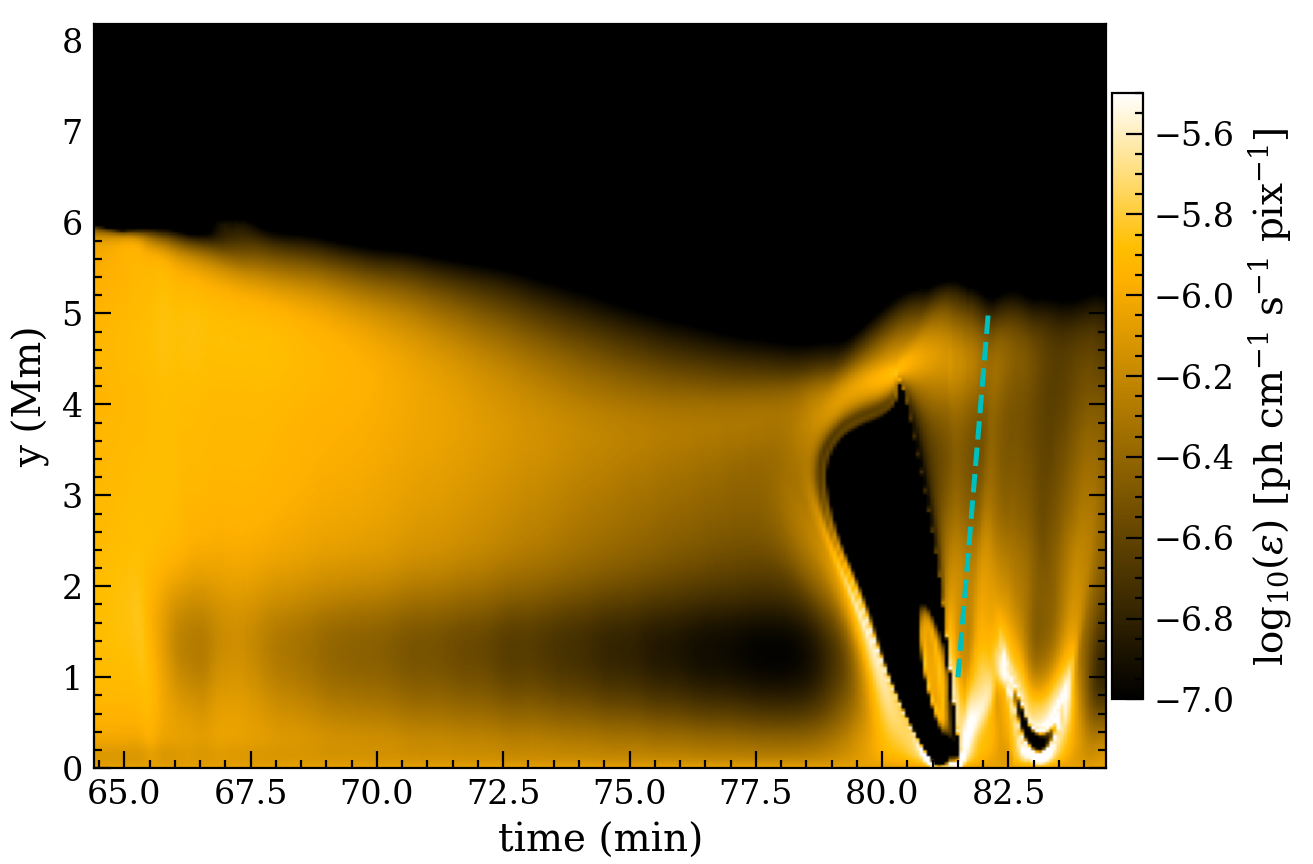}
    \caption{Top row: Synthetic emissivity maps in the SolO/HRI$_{EUV}$ 174~\AA\ channel illustrating the different phases of the coronal rain evolution. The spatial resolution has been degraded to match the HRI$_{EUV}$ pixel resolution at perihelion ($\approx 110$~km). An online animation showing the temporal evolution of the rain between $t=60.83$ and 83.01~min is available, presented at both the simulation resolution (32.6~km) and the instrument resolution (110~km). Bottom-left panel: Time-distance map of intensity, which is estimated by LOS integration along the entire vertical height, $y$. The brightening marked by the blue arrow is due to the compression of plasma leading to the heating at the downstream location of the rain clump. Bottom-right panel: Time-distance map of emissivity estimated at the vertical cut at $x=1.57$~Mm, where the slope of the brightening as marked by the dashed cyan line is $\approx 110$~km~s$^{-1}$.}
    \label{fig:HRI_rain}
\end{figure*}

\begin{figure*}[hbt!]
    \centering
    \includegraphics[width=0.32\linewidth]{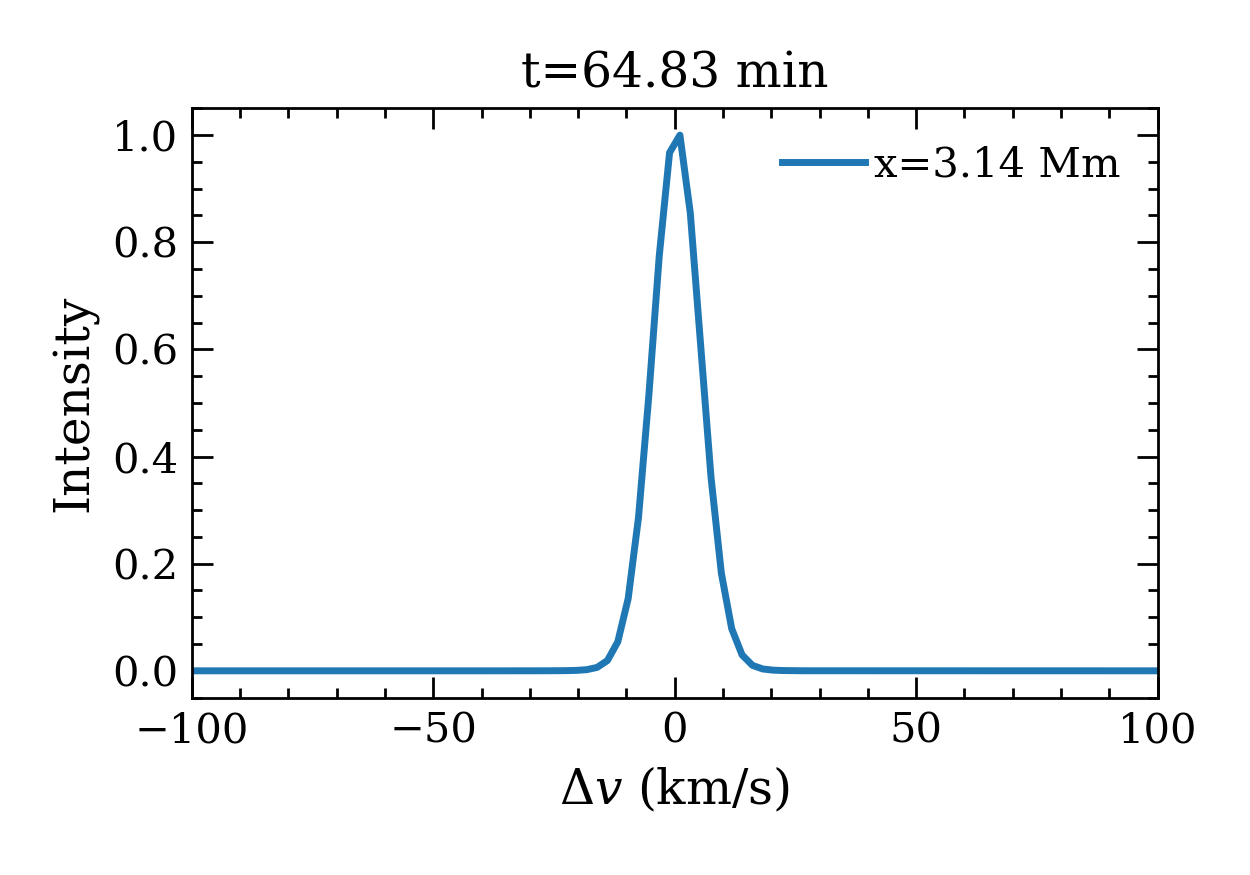}
    \includegraphics[width=0.32\linewidth]{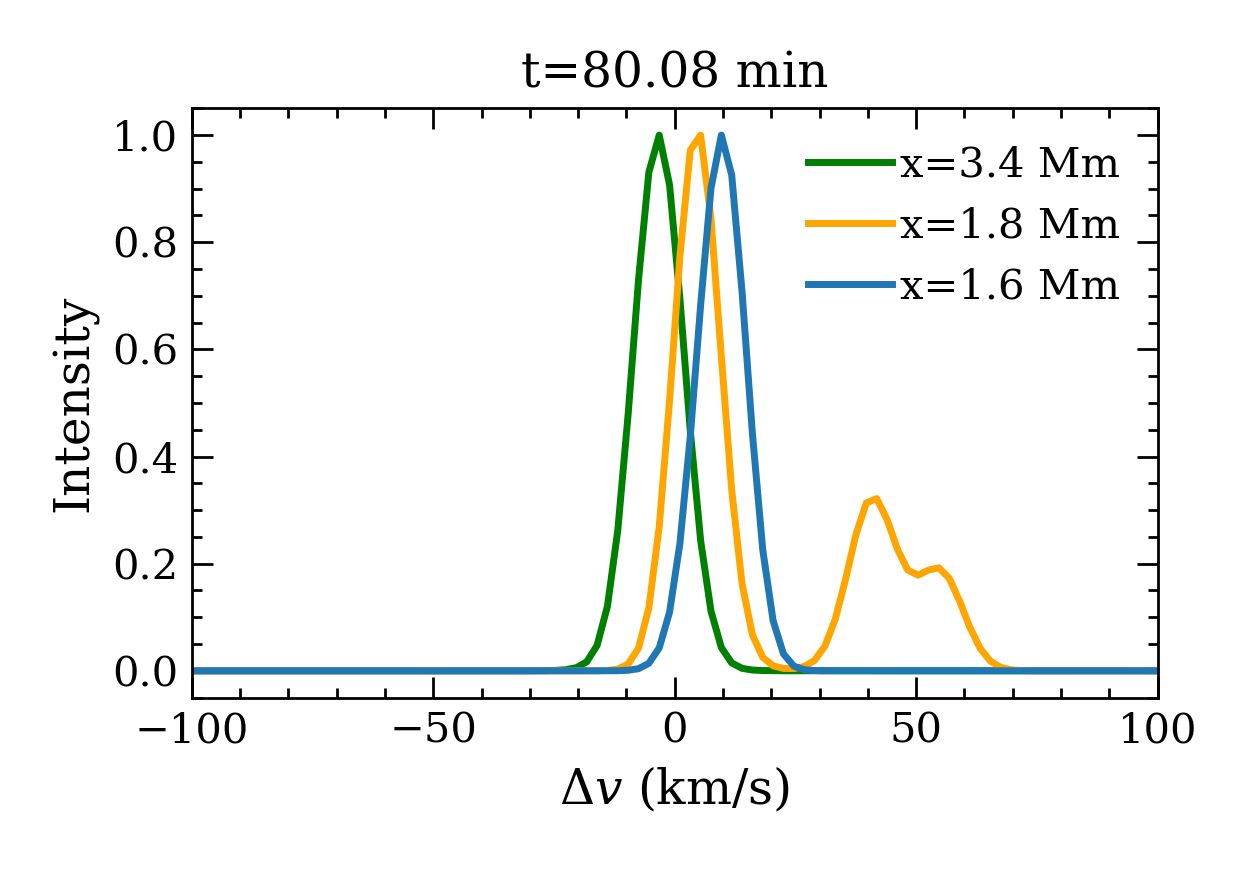}
    \includegraphics[width=0.32\linewidth]{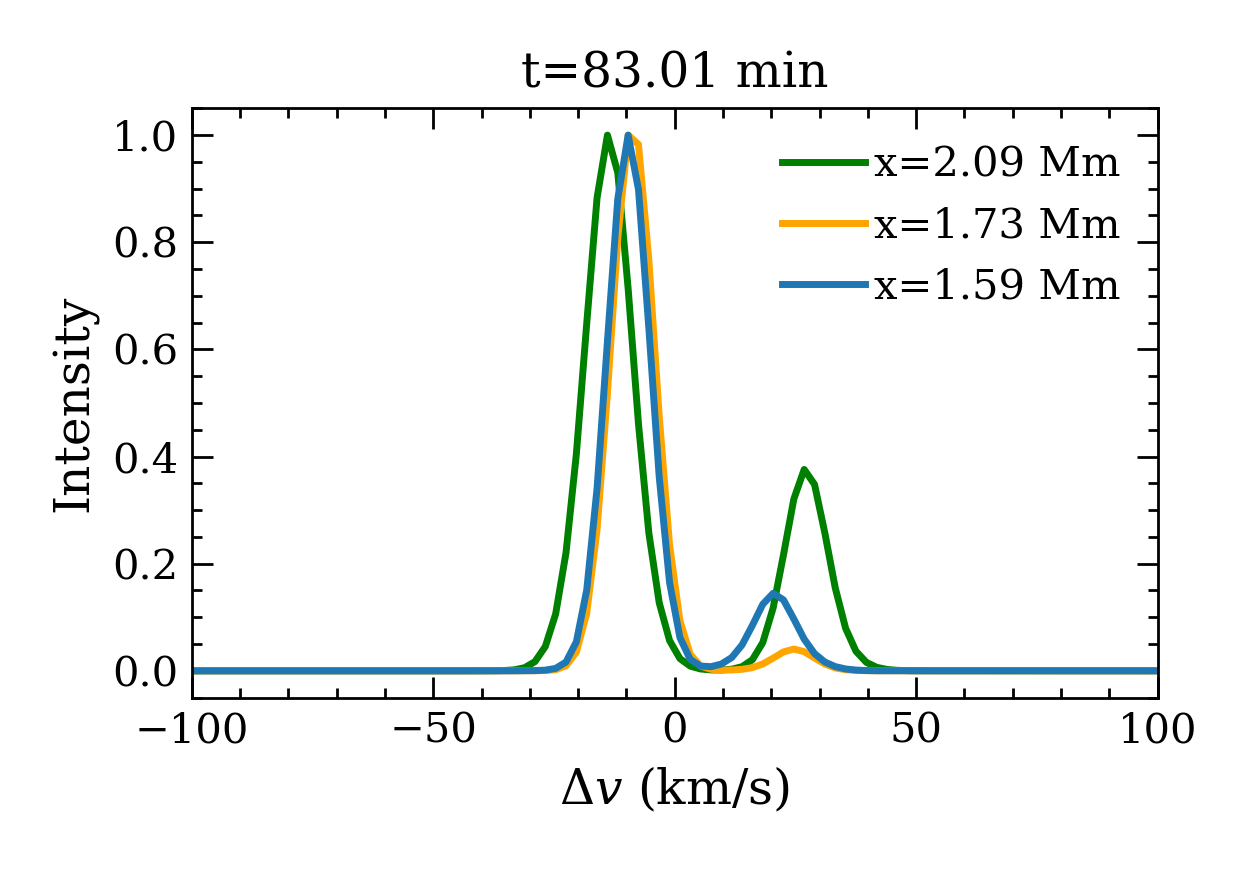}
    \caption{Synthetic spectra of Si~IV 1402.77~\AA\ line (with normalized intensities) along the vertical line of sight direction, integrating from the top boundary toward the bottom boundary of the simulation domain. The colors of the spectral profiles are chosen to correspond to those of the vertical dashed lines shown in the top row of Figure~\ref{fig:MHD_rain} at the respective times.}
    \label{fig:rain-IRIS}
\end{figure*}

\begin{figure*}[hbt!]
    \centering
    \includegraphics[width=0.8\linewidth]{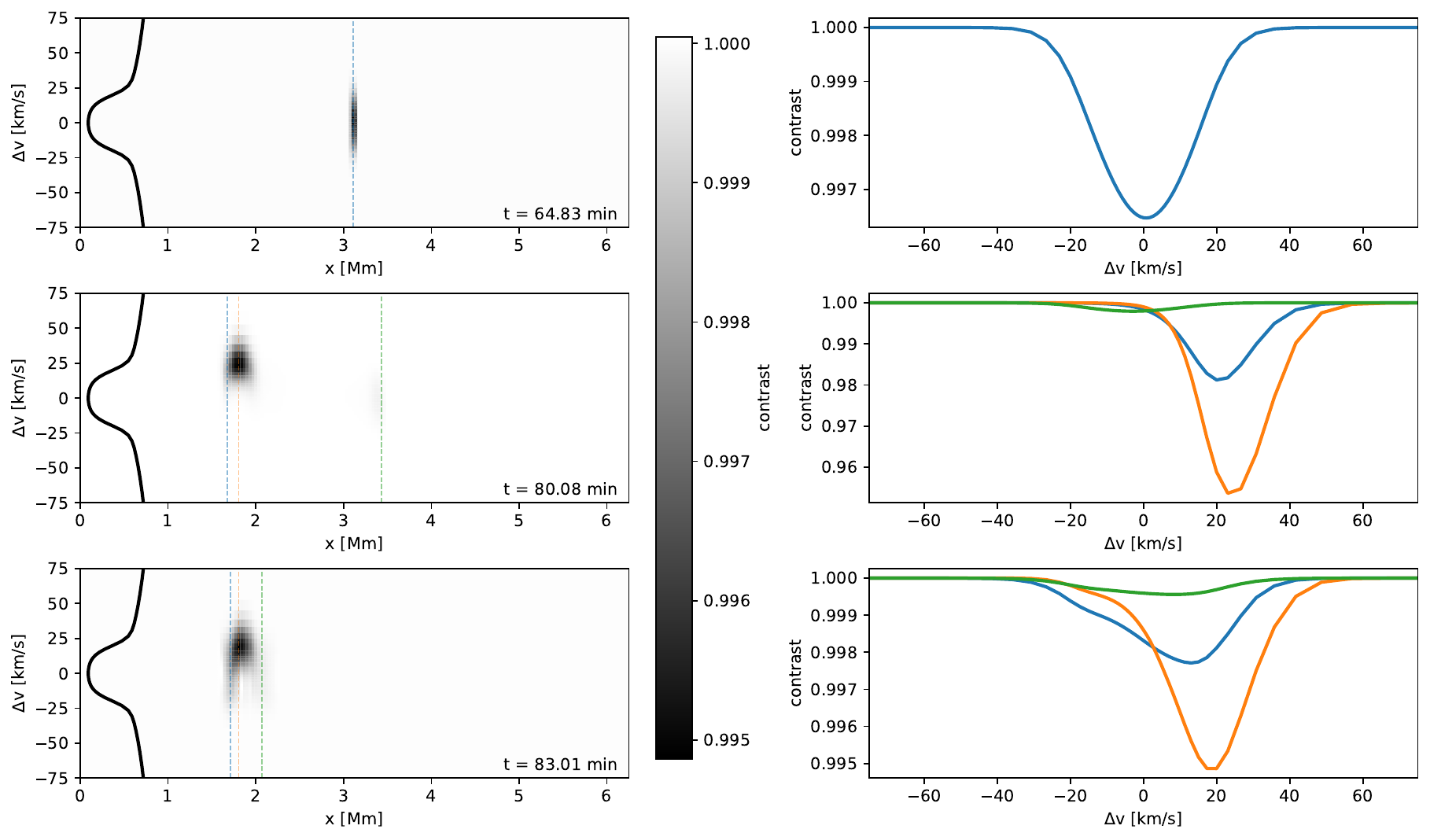}
    \caption{Contrast maps of the H$\alpha$ synthetic spectra of a filament view of three different stages of the post-flare coronal rain ($t=64.83$, 80.08, and 83.01~mins from top to bottom, respectively). Cuts along the dashed lines are shown in the right-hand panels. The black line represents a quiet H$\alpha$ profile showing the extent of the Doppler domain.}
    \label{fig:Pw_spectra}
\end{figure*}

\section{Post-flare coronal rain} \label{sec:coronal-rain}
As a consequence of the eruption of the magnetic flux ropes, localized cool-condensations are formed, which share similarities with the properties of post-flare coronal rain. We discuss the formation and evolution of these cool-condensations based on the simulation by S24, followed by the synthetic diagnostics as detailed in the following. 

\subsection{Formation and evolution}\label{sec:CR-MHD}
At a later stage following the eruptions of the flux ropes, an imbalance between heating and cooling develops at the loop apex located at $x=3.14$ and $y\approx 4$~Mm. This imbalance arises from the combined effects of field-aligned thermal conduction, steady background heating, and optically thin radiative losses, ultimately triggering thermal instability at that location at around $t=64$~min. A detailed account of the onset and evolution of the thermal instability leading to the formation of the cool-condensation site at the loop top is provided in S24. The temperature and $v_y$ distributions illustrating the formation and evolution of these cool structures are presented in the top and bottom rows of Figure~\ref{fig:MHD_rain} respectively for three different stages. The top-left panel shows the formation of the cool material at the loop apex at $t=64.83$~min. The top-middle panel corresponds to $t=80.08$~min, when the cool plasma slides downward along the loop under the influence of gravity. The top-right panel displays the stage at $t=83.01$~min, when the cool plasma merge at the lower boundary. These cool sites correspond to a temperature of $\sim 10$~kK and density of $\gtrsim 10^{-14}$~g~cm$^{-3}$ (not shown here; see S24), sharing similarities with the thermodynamical properties of post flare coronal rain. The corresponding LOS velocity ($v_y$) distributions along the height ($y$) are shown in the bottom row, where the different line colors are chosen to correspond to the vertical dashed lines overplotted on the maps in the top row of the same figure at each time. At $t=64.83$~min, the cool(-condensed) blob at $y\approx4$~Mm remains nearly static, whereas at the later stage at $t=80$~min, the downward motions of different parts of the cool blobs at $y\lesssim 5$~Mm are evident. The upward velocity appears in the green curve is due to the fragmentation of the rain clump at its trailing part. At $t=83.01$~min, we also observe upward velocities in different parts of the merged blob. This arises from the rebound of the blob upon reaching the lower boundary, where a line-tied boundary condition is imposed (see S24). We note that the denser (and cooler) parts of the rain are the ones with faster down flowing speeds, in agreement with the effects of compression and momentum transfer with the hot downstream plasma ahead of the rain \citep{Oliver_2014ApJ...784...21O, MartinezGomez_2020AA...634A..36M, Hillier_2025AA...696A.231H}. In the current work, we specifically focus on the synthetic counterparts, emphasizing the signatures of the post-flare rain in UV, EUV and optical bands, as described in the following.

\subsection{Synthetic observations in EUV and UV}\label{sec:syn_obs_EUV}
The synthetic emissivity maps in SolO/HRI$_{EUV}$~174~\AA\ during the post-flare coronal rain phase are shown in the top row of Figure~\ref{fig:HRI_rain}. Here, we degrade the synthetic maps by applying a Gaussian convolution, reducing the spatial resolution from the native simulation resolution to 110 km per pixel, which is comparable to the instrument resolution of the SolO/HRI$_{\mathrm{EUV}}$ during its perihelion phase. At $t=64.83$~min, a dark feature is visible at $x\approx3.12$~Mm and a height of $y\approx4$~Mm, embedded within a brighter surrounding region. This dark feature corresponds to a plasma density of $\sim10^{-12}$~g~cm$^{-3}$, approximately two orders of magnitude higher than the ambient coronal density. However, its temperature is only $\sim9000$~K, which is substantially lower than the peak formation temperature of the HRI$_{EUV}$~174~\AA\ passband ($\sim0.9$~MK). Consequently, the cool condensed plasma appears as a dark clump, indicated by the green arrow. The characteristic size of the clump is approximately 500~km in both spatial directions, in good agreement with the observations of \cite{Antolin:2023}, who reported rain-clump widths of $500\pm200$~km. At later stages, namely at $t=80$ and 83~min, as the rain clump continues its downward motion and subsequently rebounds from the lower boundary, the region beneath and surrounding the clump becomes brighter, as indicated by the red arrows in the top-middle and top-right panels, respectively. These brightenings are associated with plasma compression caused by the downward motion and rebound of the rain clumps, leading to localized heating. Such behavior is also found in the observations by \cite{Antolin:2023} and \citet{Wachira:2026}, where the local brightenings in the intensity maps are present in a small compression region immediately ahead of the rain and at the impact location of the rain clumps in the transition region. The bottom-left panel presents a time-distance intensity map covering the coronal rain episode from $t=64$ to 83~min. The intensity is obtained by integrating the emissivity along the LOS depth over the full height range, $y=0$-25~Mm, at each $x$ locations. The brightenings induced by the falling and the rebound rain clumps are also evident in the time-distance map and is marked by the blue arrow. The accompanying animation further demonstrates that this bright feature remains consistently located downstream of the falling rain clump throughout its evolution. The bottom-right panel shows the time--distance map of the emissivity along the vertical cut at $x=1.57$~Mm. The brightening traced by the cyan dashed line represents an upward-propagating disturbance with a speed of $\approx 110$~km~s$^{-1}$. This feature is consistent with the observation reported by \cite{Antolin:2023}, where upward-propagating disturbances with speeds of approximately 130~km~s$^{-1}$ were interpreted as the coronal response in the form of upward-propagating slow mode shocks. Slower speeds of $50-85$~km~s$^{-1}$ were reported in that work and in \citet{Wachira:2026} and were interpreted as hot upflows. Such slower speeds are missing from our model, probably due to the absence of a chromosphere. Our results therefore reinforce the observational interpretation. 

Figure~\ref{fig:rain-IRIS} represents the synthetic spectral line profiles for Si~IV~1402.77~\AA\ line at $t=64.83$, 80.08, and 83.01~min shown from the left to right panels, respectively. Here, we take the LOS integration from the top boundary towards the bottom boundary at different cuts at $x$ locations as shown in the corresponding legends that passes through the different regions of the coronal rain sites. These locations are consistent with the vertical dashed lines (with the same colors) as shown in the top panel of Figure~\ref{fig:MHD_rain}. At $t=64.83$~min, the profile peak does not have any noticeable Doppler shift from its line core, representing a nearly-static rain blob. At $t=80.08$~min, the profile for $x=1.8$~Mm (orange curve) shows a multi-modal distribution, showing a maximum red shift of $\approx 50$~km~s$^{-1}$ representing the downward motion of the rain. However, the blue ($x=1.6$~Mm) and the green ($x=3.4$~Mm) curves show the red and blue shifts of $\lesssim 10$~km~s$^{-1}$ respectively, which are (nearly) consistent with the LOS velocity profiles shown in the bottom-middle panel of Figure~\ref{fig:MHD_rain}. The presence of the blue shift (positive $v_y$) in the green curve corresponds to the trailing part of the coronal rain during the fragmentation stage of the rain clump. At $t=83.01$~min, we notice the bimodal distributions of all the three spectral profiles, with both blue and red shifts. The blue shifts here correspond to the rebound of the rain blob from the bottom boundary. The green curve shows the maximum red shift of $\approx 30$~km~s$^{-1}$, and a blue shift of $\approx20$~km~s$^{-1}$, while the blue and the orange curves have slightly lower Doppler shifts in the red and blue wings. These values are also consistent in the regime of the LOS velocity profiles in the bottom-right panel of Figure~\ref{fig:MHD_rain}. Although we suspect that the line-tying conditions significantly reduce these speeds, the obtained values are within the values reported in the same SJI passband by \cite{Sahin:2023}.

\subsection{Synthetic observations in H$\alpha$}\label{sec:syn_obs_optical}
The forward modeling of prominence spectra in optically thick lines has gained renewed attention following the work of \citet{Jenkins2023}, who used the \texttt{Lightweaver} code \citep{Osborne2021} to address the radiation trapping problem that affects other synthesis codes when modeling prominence spectra \citep{Heinzel2025}. This work formed the basis of the dedicated \texttt{Promweaver} code \citep{Promweaver030}, which handles the boundary conditions required for synthesizing prominence spectra in optically thick lines while working with stratified 1D atmospheres and optionally conserving charge and pressure \citep{Jercic2024}. These developments have made the synthesis of such lines more accessible and enabled more direct comparisons with observations \citep[e.g. ][]{Pietrow2024, Snow2025}.

More specifically, \texttt{Promweaver} employs 1.5D solution in which a limb-darkened radiation field is sampled as the boundary condition, rather than assuming a traditional plane-parallel surface \citep{Jenkins2023}. This approach allows the code to self-consistently account for Doppler shifts relative to the solar surface \citep[][]{Peat2024}. It is to be noted that only H and Ca are considered as elements in non-LTE, and that a 5-level plus continuum model is used for both atoms, which are the default atoms taken from the RH code \citep{Uitenbroek2001}. Furthermore, as the MHD model (S24) uses a fully ionized medium, we use tables from \cite{Heinzel:2015} for the height of 10\,Mm as an initial guess for the ionization degree, which is iterated by the code for charge neutrality with the final non-LTE state. Lastly, the turbulent velocity is calculated according to equation (13)-(16) of \cite{Heinzel:2001}.

Our current MHD model does not include the chromosphere, therefore the semi-empirical FAL-C atmosphere \citep{Fontenla1993} is adopted. In \texttt{Promweaver}, FAL-C is included as part of the boundary conditions to prevent the nonphysical interaction between it and the filament, which is unavoidable in a 1.5D approximation. This is also known as radiation trapping \citep{Paletou1993} and its influence and implementation in \texttt{Promweaver} are described in detail in \cite{Jenkins2023}. We then used the standard settings for the formal solver and synthesized the simulation at full (32.6~km pixel size) resolution along the 6.28~Mm domain in so-called filament mode, where the integration direction begins above the simulation and proceeds downward toward the solar surface, thus emulating a filament-viewing geometry. The coronal rain clumps in our simulation exhibit temperatures of around $10000$~K, reaching a minimum value of around 9000~K. Since the H$\alpha$ line has a peak formation temperature near $10^4$~K, this motivates us to select this line for the spectral synthesis in order to highlight the absorption features associated with the coronal rain sites. This is, in fact, one of the key spectral lines widely used to identify coronal rain signatures in observations (e.g., \citealt{Antolin_etal_2012SoPh..280..457A, Froment:2020, Schmidt:2025}, and references therein).

Figure~\ref{fig:Pw_spectra} shows synthetic spectra of H$\alpha$ corresponding to the three snapshots whose temperature distribution is shown in Fig.~\ref{fig:MHD_rain} ($t=64.83$, 80.08, and 83.01~mins, from top to bottom respectively). The left panels show contrast profiles, where each spectrum is normalized to the first in the domain to emphasize differences from a typical H$\alpha$ profile. These correspond to the filament view, with the horizontal axis representing the simulation's $x$-coordinate and the vertical axis showing Doppler-shifted velocities. A typical H$\alpha$ profile is overplotted in black to better show the magnitude of the Doppler shifts. The normalized spectra along the dashed lines are shown in the right panels in their respective colors. These are consistent with the vertical dashed lines shown in the top row of Figure~\ref{fig:MHD_rain}, along with their corresponding LOS velocity profiles displayed in the bottom row. The formation height of the H$\alpha$ line corresponds to the locations where the slits intersect the coronal rain blobs, whose temperatures are close to the peak formation temperature of the H$\alpha$ line ($\approx 10$~kK). We note that the strongest absorption profiles correspond to the faster down flowing regions, reflecting the densest (and coolest) regions of the rain. The top row of Figure~\ref{fig:Pw_spectra} shows the moment of the rain blob formation, and we see the cold (and dense) structure from Fig.~\ref{fig:MHD_rain} appearing as increased absorption around 3.14~Mm. At this moment, the structure is predominantly static, and we see it centered around 0\,km\,s$^{-1}$. The following row, shows H$\alpha$ at 80.08\,min. We see the condensed structure extended along $x$ and now significantly red-shifted. In the right panel of the middle row, we see the Doppler shifted value is $\approx23$\,km\,s$^{-1}$. Furthermore, different parts of the falling coronal rain blob have different absorptions as they cut through different parts of the extended structure. The central part of the structure (orange cut) shows the strongest absorption ($\sim4$\%) in comparison to the typical background profile. Integrating the H$\alpha$ line-core opacity along each LOS results in a maximum opacity of $\tau \approx 0.1$, meaning that the medium is optically thin in each of the snapshots shown here. The strongest absorption at $t=80.08$\,min is directly related to the fact that the LOS is passing through more of the dense material (see the top-middle panel of Fig.~\ref{fig:MHD_rain}). The bottom row at $t=83.01$\,min catches the rain already at the bottom of the domain, significantly compressed but still falling. As a result, the bottom-most right panel in Fig.~\ref{fig:Pw_spectra} shows a maximal red-shift of $\sim19$\,km\,s$^{-1}$ and a significant decrease in the total absorption as the dense material shrunk.

\section{Summary and Discussion} \label{sec:summary}
In this work, we present synthetic observations that enable diagnostics of a flux rope-trapped mini-prominence (or mini-filament) eruption and the associated post-flare coronal rain, based on the 2.5D MHD model of S24. We perform forward modeling under the optically thin approximation in EUV and UV channels corresponding to the SolO/HRI$_{EUV}$ 174~\AA\ and IRIS/SJI 1400~\AA\ passbands, respectively. In addition, we carry out spectral synthesis in Si IV 1402.77~\AA\ line using optical-thin approximation, and use radiative transfer treatment taking into account the non-LTE approximation for the H$\alpha$ line.

The SolO/HRI$_{EUV}$ 174~\AA\ synthetic maps in Figure~\ref{fig:HRI} reveal brightening associated with the flux rope during both the formation stage at $t=12.59$~min and the eruptive phase at $t=13.38$~min. The rim-like brightening around a height of $y=19$~Mm in the bottom-left panel (simulation resolution) represents a flux rope of size $\sim$~few Mm, enclosing a darker region within the rim. This region appears fainter, yet remains discernible, in the corresponding image in the bottom-right panel with degraded resolution comparable to that of SolO/HRI$_{EUV}$.

The dark core region inside the erupting flux rope in the SolO/HRI$_{EUV}$ maps corresponds to a bright region in the IRIS/SJI 1400~\AA\ synthetic maps shown in Figure~\ref{fig:IRIS}. This localized brightening is caused by the presence of cool plasma with a temperature of around 40~kK, as the peak formation temperature of the IRIS/SJI 1400~\AA\ response function is around 80~kK. The associated animations in Figures~\ref{fig:HRI} and \ref{fig:IRIS} demonstrate that the bright region in the IRIS/SJI 1400~\AA\ images and the dark region inside the flux rope of the SolO/HRI$_{EUV}$ 174~\AA\ maps are co-spatial and remain trapped within the erupting flux rope. This phenomenon corresponds to a mini-prominence eruption that is confined within the dip of the flux rope. The time evolution shown in the bottom panel of Figure~\ref{fig:IRIS} illustrates the growth of the size of the mini-prominence starting from $t \approx 12.7$~min to $13.4$~min, reaching a maximum area of approximately $0.25$~Mm$^2$. Thereafter, the area gradually decreases as the prominence material becomes progressively confined within the dip of the flux rope until $t \approx 13.9$~min, after which it exits the simulation domain. The formation of this mini-prominence can be interpreted within the framework of in-situ condensation, namely the ``reconnection-condensation model'' \citep{Kaneko:2017}. In this scenario, magnetic reconnection forms a flux rope that lifts the overlying coronal plasma into a closed magnetic structure, thereby creating favorable conditions for cool-condensation. The reconfiguration of the magnetic topology thus acts as the primary trigger for plasma cooling and mass accumulation.

The spectral synthesis of the Si IV 1402.77~\AA\ line (top-left panel of Figure~\ref{fig:prominence-spectra}) indicates that the maximum eruption speed of the mini-filament material reaches approximately 250 km~s$^{-1}$, which has a reasonable agreement with observational results reported by \cite{Chen-minifilament:2020, Devi:2021} (and references therein). Furthermore, the synthetic spectra reveal the presence of two predominant upward velocity components of 220~km~s$^{-1}$ and 250 km~s$^{-1}$, indicating a relative motion of the different parts of the mini-filament material during the eruptive stage at $t=13.38$~min. This result motivates future observational campaigns to investigate this scenario in a greater detail for mini-filament eruptions.

In an advanced stage of the evolution, we observe the formation of a cool, condensed blob at the loop apex of the post-flare arcade at $t=64$~min, approximately $30$~min after all the flux ropes leave the simulation domain (see the top-left panel of Figure~\ref{fig:MHD_rain}). The blob gradually increases in size and subsequently slides downward along the arcade under the influence of gravity (see the top-middle panel of Figure~\ref{fig:MHD_rain}), eventually merging at the bottom boundary at $t=83$~min (see the top-right panel of Figure~\ref{fig:MHD_rain}). These timescales are consistent with observations of a C8.2-class post-flare coronal rain event, where the time delay between the formation of flare ribbons and the first appearance of coronal rain in post-flare loops is $\approx 26$~min \citep{2016:scullion}. The observational evidence indicating that coronal rain forms near loop-top regions and subsequently falls along magnetic field lines is in good agreement with our simulation results. The condensation timescale also shows reasonable agreement with observations of an X2.1-class post-flare coronal rain event, in which the time interval between the peak phase of the flare and the first appearance of condensation at the loop apex is $\approx 50$~min \citep{2024:Brooks}. 
 
The synthetic imaging in SolO/HRI$_{EUV}$~174~\AA\ for the coronal rain phase shows the brightening at the downstream location of the falling rain clump, and at its surrounding when it rebounds from the lower boundary (see Figure~\ref{fig:HRI_rain}). The brightenings are associated with the compression of the plasma leading to the localized heating. We further obtain upward propagating disturbances at a speed of 110~km~s$^{-1}$, which we interpret as slow mode shocks in response to the impact. This is also consistent with the observations of \cite{Antolin:2023} and \citet{Wachira:2026}, where localized intensity enhancements were detected both immediately downstream and at the surroundings of coronal rain clumps during impact. Only hot flows or a mix of flows and rebound shocks were observed in response to the rain impact. The absence of hot upflows in our simulation is likely due to the lack of a chromosphere.

The spectral synthesis in the H$\alpha$ line reveals the presence of dark regions at the coronal rain sites (see the left column of Figure~\ref{fig:Pw_spectra}), indicating enhanced absorption at the cool condensation locations relative to the background. This absorption becomes more pronounced at later stages ($t=80$ and 83~mins) compared to the initial formation stage of the rain clump at $t=64$~min. This suggests that the falling clump becomes denser and/or cooler than at its formation stage. The estimated Doppler shifts of the rain clumps indicate a redshift relative to the line core, implying downflows that reach a maximum velocity of approximately 23~km~s$^{-1}$ (see the middle-right panel of Figure~\ref{fig:Pw_spectra}). On the other hand, the spectral synthesis of the Si~IV~1402.77~\AA\ line indicates that the maximum velocity of the falling rain clump is approximately 50~km~s$^{-1}$. The difference between the velocities inferred from these two diagnostics suggests the presence of thermodynamic and dynamic substructure within the falling rain clump. These values are in reasonable agreement with observational results reported by \cite{2015:patrick, 2019:mason}, where the velocities of rain clumps are found to lie in the range between $\approx 50$--$100$~km~s$^{-1}$. The line-tying conditions in our model prevent further acceleration of the rain, as the plasma pressure beneath the falling rain blobs becomes higher than that within the coronal rain blobs (see S24), which is also similar to the results by \cite{Mackay:2010, AdroverGonzalez_2021AA...649A.142A}. Furthermore, the synthetic diagnostics suggest fragmentation of the rain blobs into components with different thermodynamics and LOS velocities during their downward motion, a region that is often challenging to infer from observations due to the (spatial and spectral) resolution limit of the telescopes.  

However, the synthetic diagnostics successfully capture the eruptive mini-prominence (or mini-prominence) trapped within the flux rope, as well as the post-flare coronal rain based on a unified MHD model by S24, one limitation of the model lies in its dimensionality. The 2.5D approximation does not allow us to investigate whether mass drainage of prominence material occurs at one of the (or both) loop footpoints during the eruption phase, which is an important aspect for understanding the stability and life cycle of prominence. Future extensions of this model should address this limitation by employing fully 3D simulation to assess this scenario. Another improvement of this model should include coupling to the lower chromosphere and photospheric layers. A localized heating prescription at the loop footpoints may also trigger TNE, thereby promoting and regulating the formation of coronal condensations, as proposed by \cite{1991:antiochos}. \cite{2022:xiaohong} demonstrated the formation of coronal rain showers driven by turbulent heating at chromospheric footpoints. The present model could therefore be advanced by extending the spatial domain from the chromosphere to the corona and incorporating localized heating at loop footpoints. Such improvements may lead to a quasi-cyclic formation of a larger number of rain clumps or rain showers.

It is worth noting that we do not introduce any readout or Poisson noise when producing the synthetic observables. Conducting analyses without noise is a standard practice, as demonstrated in studies of rotating prominences \cite{Valeriia:2023, Pietrow2024}, as well as in investigations of microphysical processes such as MHD turbulence \citep{Sen:2021, Shen:2022NatAs}. However, the extent to which the inclusion of additional instrumental effects, noise, and the presence of increased material along the LOS \citep{Ernest:2025}, or an extended integration time may smooth or smear the signatures of mini-prominence and coronal rain structures remains an important topic for future investigation.

Nevertheless, this work highlights the importance of generating synthetic observables of mini-prominence eruptions and post-flare rain in order to provide predictions for their detectability with current observing facilities. It also establishes an important link between modeling and observations of homologous eruptions, prominences (or filaments), and post-flare rain, thereby opening avenues for a deeper understanding of the mechanisms underlying these phenomena in the solar atmosphere.

\begin{acknowledgements}
We thank the anonymous referee for the useful and constructive suggestions, which have improved the manuscript considerably. SS acknowledges support by the European Research Council through the Synergy Grant \#810218 (``The Whole Sun”, ERC-2018-SyG) and the Research Council of Finland through the Centre of Excellence project, SpaceResilience (Grant \#374096). SS thankfully acknowledges the technical expertise and assistance provided by the Spanish Supercomputing Network (Red Espa\~{n}ola de Supercomputaci{\'o}n), as well as the computer resources used: the LaPalma Supercomputer, located at the Instituto de Astrof{\'i}sica de Canarias. SS acknowledges D. N{\'o}brega-Siverio for providing the response function of the SolO/HRI$_{EUV}$~174~\AA. AP is supported by the {Deut\-sche For\-schungs\-ge\-mein\-schaft, DFG\/} project number PI 2102/1-1. VJ's research was made possible by an appointment to the NASA Postdoctoral Program at the Goddard Space Flight Center, administered by Oak Ridge Associated Universities under contract with NASA.

\end{acknowledgements}

\bibliographystyle{aa} 
 \bibliography{reference}

\end{document}